\documentclass[twocolumn,twocolappendix,times,tighten,trackchanges]{aastex631} %this is default
\usepackage[version=3]{mhchem}
\usepackage{comment}
\usepackage{multirow}

\newcommand{\mum}{$\rm \mu \mathrm{m}~$}

\newcommand{\OIII}{[\textsc{Oiii}]\,}

\newcommand{\HII}{H{\sc ii}~}
\newcommand{\CII}{[C{\sc ii}]\,}
\defcitealias{Sugimura:2024}{S24}

\usepackage{color}
\usepackage{makecell}

\shorttitle{Clump-Scale Dust Attenuation in EoR Galaxies}
\shortauthors{Nakazato et al.}

\graphicspath{{./}{figures/}}
\newcommand{\figdir}{./}
\begin{document}

%\title{Spatially Resolved Dust Properties in Clumpy Galaxies at the Epoch of Reionization: Merger vs. Violent Disk Instability}
%\title{Clump-Scale, Spatially Resolved Dust Properties in EoR Galaxies from FirstLight Simulations}
\title{Clump-Scale Dust Attenuation in Epoch of Reionization Galaxies: Spatially Resolved Properties from FirstLight Simulations}

\correspondingauthor{Yurina Nakazato}
\email{ynakazato@flatironinstitute.org}
\author[0000-0002-0984-7713]{Yurina Nakazato}
\affiliation{Center for Computational Astrophysics, Flatiron Institute, 162 5th Avenue, New York, NY 10010}
\affiliation{Department of Physics, The University of Tokyo, 7-3-1 Hongo, Bunkyo, Tokyo 113-0033, Japan}
\author[0000-0002-5012-6707]{Kosei Matsumoto}
\affiliation{Sterrenkundig Observatorium Department of Physics and Astronomy Universiteit Gent, Krijgslaan 281 S9, B-9000 Gent, Belgium}
\author[0000-0002-7779-8677]{Akio K. Inoue}
\affiliation{Waseda Research Institute for Science and Engineering, Faculty of Science and Engineering, Waseda University, 3-4-1 Okubo, Shinjuku, Tokyo 169-8555, Japan}
\affiliation{Department of Physics, School of Advanced Science and Engineering, Faculty of Science and Engineering, Waseda University, 3-4-1, Okubo, Shinjuku, Tokyo 169-8555, Japan}
\author[0000-0002-8680-248X]{Daniel Ceverino}
\affiliation{Departamento de Fisica Teorica and CIAFF, Facultad de Ciencias, Universidad Autonoma de Madrid, Cantoblanco, 28099 Madrid, Spain}
\affiliation{CIAFF, Facultad de Ciencias, Universidad Autonoma de Madrid, 28049 Madrid, Spain}
\author[0000-0003-3127-5982]{Takashi Hosokawa}
\affiliation{Department of Physics, Graduate School of Science, Kyoto University, Sakyo, Kyoto 606-8502, Japan}
\author[0000-0003-3467-6079]{Daisuke Toyouchi}
\affiliation{Theoretical Astrophysics, Department of Earth and Space Science, The University of Osaka, 1-1 Machikaneyama, Toyonaka, Osaka
560-0043, Japan}

\begin{abstract}
    Understanding dust attenuation in galaxies at both integrated and spatially resolved scales is fundamental for accurately determining the physical properties of galaxies. Recent high-spatial-resolution observations with ALMA and JWST enable investigations of spatially resolved properties in high-redshift galaxies ($z \gtrsim 6$), but spatial variations in dust observables remain poorly constrained. We use cosmological zoom-in simulations combined with postprocessing dust radiative transfer calculations for 376 clumpy galaxies at $z=6$-$9$ with stellar masses of $M_* \gtrsim 10^9 \, M_\odot$. For each system, we investigate dust attenuation and reemission properties for three components: system-integrated, individual clumps, and diffuse regions. We find that system-integrated attenuation curves are grayer than the Calzetti curve, even when assuming MW- or SMC-type dust. Attenuation curves of individual clumps are even grayer, while diffuse regions exhibit steeper curves owing to enhanced scattering in optically thin environments. Since the effects of optical depth and dust–star geometry are intrinsically degenerate in attenuation curves, we introduce a toy model based on the IRX–$\Delta\beta$ plane, where $\Delta\beta$ denotes the difference between attenuated and intrinsic UV slopes. Applying this framework, we find that clumps have dust column densities approximately an order of magnitude higher than system-integrated values and exhibit co-spatial or dust-extended geometries. In contrast, system-integrated attenuation reflects star-extended geometries driven by contributions from optically thin diffuse regions. We apply this framework to REBELS-IFU galaxies at $z \sim 7$ and find good agreement with our simulation predictions.
\end{abstract}
\section{Introduction} \label{sec:introduction}
Astrophysical dust is a key component of the interstellar medium (ISM) in galaxies. The composition and size distribution of dust grains determine the wavelength dependence of absorption and scattering, which, combined with the spatial distribution of dust relative to radiation sources (e.g., stars), determines the overall attenuation. The absorbed radiation is reradiated as infrared (IR) radiation. This dust attenuation and reemission shape the observed spectral energy distributions (SEDs) of galaxies.

Understanding dust attenuation and reemission is crucial for accurately measuring fundamental galaxy properties. The shape of the dust attenuation curve, typically characterized by the ratio of attenuation at different wavelengths (e.g., $A_{\rm UV}/A_{\rm V}$, where $A_\lambda \equiv 2.5 \log_{10} (f_\lambda^{\rm int}/f_\lambda^{\rm obs})$), is essential for correcting observed fluxes and deriving intrinsic properties such as stellar masses and star formation rates (SFRs). The ultraviolet (UV) continuum slope ($\beta_{\rm UV}$), defined by $f_\lambda \propto \lambda^{\beta_{\rm UV}}$ \citep{Calzetti:1994}, is widely used as an indicator of dust attenuation. Absorbed UV and optical photons are reemitted in the IR, and the ratio of total IR to UV luminosity (infrared excess, IRX $\equiv L_{\rm IR}/L_{\rm UV}$) provides a measure of dust obscuration, often evaluated through the IRX--$\beta_{\rm UV}$ relation \citep{Calzetti:1994, Meurer:1999, Reddy:2006, Overzier:2011}. 

The advent of James Webb Space Telescope (JWST) and the Atacama Large Millimeter/submillimeter Array (ALMA) has revolutionized our ability to study dust properties during the Epoch of Reionization (EoR; $z\simeq 6$--9). JWST spectroscopic observations provide rest-frame UV-to-optical continua, enabling direct measurements of attenuation curves for individual galaxies \citep[e.g.,][]{Markov:2023, Markov:2025, Shivaei:2025, Fisher:2025}. These observations reveal that attenuation curves at the EoR are systematically flatter than the empirical Calzetti curve\citep{Calzetti:2000}, while UV bump detections provide constraints on dust composition \citep[e.g.,][]{Witstok:2023}. Complementary ALMA observations in the far-IR measure dust temperatures and total IR luminosities \citep[e.g.,][]{Inoue:2020, Inami:2022, Sommovigo:2022, Mitsuhashi:2024, Algera:2024, Bowler:2024}, showing increasing dust temperatures at higher redshifts and deviations from the canonical IRX--$\beta_{\rm UV}$ relation in \citet{Meurer:1999}.

Beyond galaxy-integrated measurements, spatially resolved observations are beginning to reveal significant variations in dust attenuation properties within individual high-redshift galaxies. High-resolution JWST/NIRCam photometry enables pixel-by-pixel SED fitting to generate maps of $A_{\rm V}$ and $\beta_{\rm UV}$ \citep[e.g.,][]{Abdurrouf:2023, Gimenez-Arteaga:2023, Gimenez-Arteaga:2024, Sun:2024, Tanaka:2024, Lines:2025, Markov:2026}, while NIRSpec IFU observations allow construction of Balmer decrement maps through H$\alpha$/H$\beta$ ratios \citep[e.g.,][]{Parlanti:2025, Ishikawa:2025, Mawatari:2026}. These observations commonly reveal clumpy morphologies in EoR galaxies \citep[e.g.,][]{Chen:2023, Hainline:2024_clumpy, Shibuya:2025, Lines:2025, Harikane:2025_clumpy}, and ongoing programs (e.g., RIOJA:GO-1840, GA-NIFS:GTO, REBELS-IFU:GO-1626) are beginning to resolve physical properties of individual clumps and diffuse components. These advances in resolving sub-galactic structure enable us to address fundamental questions about spatial variations in dust attenuation/reemission properties and the physical processes that drive them.

On the theoretical side, cosmological simulations combined with dust radiative transfer (RT) have made significant progress in reproducing observed dust attenuation/reemission properties of high-redshift galaxies. Previous studies have investigated system-integrated properties such as attenuation curves \citep{Narayanan:2018_attenuation_curve}, dust temperatures \citep{Ma:2019, Liang:2019, Vijayan:2019}, and IRX--$\beta_{\rm UV}$ relations \citep{Narayanan:2018_IRX_beta,Behrens:2018,Vijayan:2019, Liang:2021_IRX_beta,  Pallottini:2022,  Mushtaq:2023}. 
%However, these studies have focused exclusively on galaxy-integrated measurements and have not explored how dust properties vary spatially within galaxies, specifically how individual star-forming regions (clumps) differ from diffuse components and system-integrated values. 
Most of these studies have focused on galaxy-integrated measurements, and while some have noted the role of patchy dust-star geometries in shaping attenuation properties \citep{Behrens:2018, Ma:2019}, a systematic and quantitative component-by-component analysis comparing individual star-forming clumps, diffuse components, and system-integrated values has not been performed.
This gap becomes critical as spatially resolved observations become increasingly common. If dust attenuation of starlight differs significantly between clumps and their host galaxies' integrated values, analyses that assume uniform attenuation may yield systematically biased estimates of stellar masses, ages, and SFRs.
%If dust properties differ significantly between clumps and their host galaxies' integrated values, analyses that assume uniform properties across a galaxy may yield systematically biased estimates of stellar masses, ages, and star formation rates.

In this paper, we investigate how dust attenuation ($\beta_{\rm UV}$, $A_{\rm V}$, attenuation curves) and reemission properties (IRX, $T_{\rm d}$) vary spatially within high-redshift clumpy galaxies using cosmological zoom-in simulations. We analyze massive galaxies ($M_* \sim 10^{8-10}\, M_\odot$) at $z = 6$--9, combined with postprocessing dust radiative transfer calculations using SKIRT \citep{Baes:2011}. We identify 376 clumpy systems containing 1059 individual star-forming clumps ($R \gtrsim 100\, {\rm pc}$, ${\rm SFR} \gtrsim 1\, M_\odot\, {\rm yr^{-1}}$) based on SFR surface density. For each system, we analyze dust properties separately for clumps, diffuse regions, and system-integrated values. To disentangle the effects of optical depth and dust-star geometry on the observed variations, we develop a toy model that utilizes the IRX-$\Delta\beta$ plane ($\Delta\beta \equiv \beta_{\rm UV} - \beta_{\rm UV,0}$, the difference between attenuated and intrinsic UV slopes) with two key parameters: dust optical depth (i.e., dust column density) and the dust-to-star scale-height ratio.

This paper is organized as follows. Section~\ref{sec:method} describes our simulation setup, dust radiative transfer calculations, and clump identification method. Section~\ref{sec:result} presents spatially resolved dust properties for an example galaxy and statistical results for all clumpy systems. Section~\ref{sec:discussion} develops the toy model to interpret observed variations and discusses implications for observations. We summarize our findings in Section~\ref{sec:conclusion}.

\section{Method} \label{sec:method}
\subsection{Cosmological Zoom-in Simulations: FirstLight}  \label{subsec:FirstLight}
We use the cosmological zoom-in simulation suite FirstLight. The simulations used in this work were first presented in \citet{Ceverino:2017} and have already been used to study several high-redshift topics, including galaxy morphologies \citep{Ceverino:2021, Nakazato:2024, Dome:2024}, star formation histories \citep{Ceverino:2018, Ceverino:2024}, chemical evolution \citep{Langan:2020, Nakazato:2023}, UV/IR continuum emission \citep{Ceverino:2019, Mushtaq:2023}, and size evolution \citep[]{Cataldi:2026, Ceverino:2026}. 

We use the simulation code ART, which employs adaptive mesh refinement \citep[AMR;][]{Kravtsov:1997, Kravtsov:2003, Ceverino:2009}, with a parent cosmological box of side length 40 comoving Mpc $h^{-1}$. Initial conditions were generated with MUSIC \citep{Hahn_Abel:2011} at $z=150$ using a Planck cosmology \citep{Planck:2014}. We select all dark matter halos with a maximum circular velocity greater than $V_{\rm circ} = 178\, {\rm km\,s^{-1}}$ at $z=5$, resulting in a sample of 62 halos. For each halo, we perform zoom-in simulations with a dark matter particle mass of $8 \times 10^4\, M_\odot$ and a minimum stellar particle mass of $10^3\, M_\odot$. All simulations were evolved down to $z = 5.25$. The smallest cell size achieved is $\Delta x_{\rm min}=31\, {\rm pc}\,(7/(1+z))$. To prevent artificial fragmentation at the smallest cell scale, a pressure floor is imposed to ensure that the Jeans length is always resolved by at least seven cells \citep[see also Appendix~B of \citealp{Nakazato:2024}]{Ceverino:2010}.

Stars form in high-density ($\rho > \rho_{\rm th}$) and low-temperature ($T < T_{\rm th}$) gas cells. In our simulations, we adopt thresholds of $\rho_{\rm th} = 0.035\, M_\odot\, {\rm pc^{-3}}$ ($n_{\rm th} = 1\,{\rm cm^{-3}}$) and $T_{\rm th} = 10^4\, {\rm K}$. We adopt a stochastic model in which the star formation probability scales with the gas free-fall time \citep{Ceverino:2009}. This prescription ensures that higher-density gas cells are more likely to form stars, resulting in the Kennicutt–Schmidt relation \citep{Kennicutt:1998}. Our simulations follow radiative cooling by atomic hydrogen and helium, metal ions and atoms, and molecular hydrogen calculated using \textsc{cloudy} \citep{Ferland:1998}. We also include photoionization heating by the redshift-dependent UV background with partial self-shielding \citep{Haardt_Madau:1996}.

In the FirstLight simulations, we incorporate the following stellar feedback mechanisms: supernovae (SNe) and stellar winds, radiation pressure on dust grains that absorb UV photons, and IR radiation pressure via a subgrid model. The first two are implemented as thermal pressure terms with a constant heating rate for 40 Myr, whereas the latter two are implemented as non-thermal pressure terms \citep{Ceverino:2014}. In addition to thermal feedback, we include the injection of momentum from the (unresolved) expansion of gaseous shells driven by SNe and stellar winds \citep{Ostriker_Shetty:2011}. Further details can be found in \citet{Ceverino:2017}.

The 62 galaxies in our sample have stellar masses greater than $\sim 10^{10}\, M_\odot$ at $z=5$. The maximum resolution in the zoom-in hydrodynamical simulations, particularly in dense clumps, enables us to resolve gas densities of $\sim 10^3\, {\rm cm^{-3}}$ with temperatures of $\sim 300\, {\rm K}$. We stored a total of 66 snapshots from $z=9.5$ to $z=5.25$ for each galaxy, with spacing in the cosmic expansion parameter of $\Delta a = 0.001$, corresponding to 7-10 Myr. This frequent output cadence is sufficient to follow the dynamics during galaxy mergers and the subsequent clumpy phase, which occur within a dynamical timescale \citep{Nakazato:2024}. 

 \subsection{Postprocessing Radiative Transfer} \label{subsec:skirt}
Dust attenuation of all simulated galaxies (62 galaxies, 4092 snapshots) is calculated using the postprocessing radiative transfer code SKIRT \citep[version 9, ][]{Baes:2011, Camps_Baes:2020}. SKIRT is an open-source, three-dimensional Monte Carlo dust RT code that self-consistently calculates dust temperatures and reemission by accounting for both absorption and anisotropic scattering processes.
To perform the RT calculations, we specify the stellar emission properties (Section \ref{subsubsec:stellar_emission}), dust spatial distribution (Section \ref{subsubsec:dust_spatial_distribution}), and adopted dust composition and grain-size distributions (Section \ref{subsubsec:dust_graine_properties}) as input parameters. For each simulation snapshot, we extract a cubic region with a side length of 10 kpc centered on the main halo and perform radiative transfer calculations on the extracted volume.

\subsubsection{Stellar Emission} \label{subsubsec:stellar_emission}
%The spatial distribution of stellar emission is determined by the positions of stellar particles in our hydrodynamical simulations. \textsc{SKIRT} supports directly importing data files that contain the necessary information on these stellar particles.
We construct the SEDs of individual stellar particles using binary stellar population models from \textsc{BPASS} \citep{Eldridge:2017}, based on the mass, age, and metallicity of each stellar particle. Each stellar particle is treated as a single starburst. Among the \textsc{BPASS} models, we adopt a Chabrier initial mass function (IMF) with an upper mass limit of $300\, M_\odot$.

For stellar particles younger than 15 Myr, we reassign their ages following \citet{Nakazato:2023}. This adjustment is needed because our simulations form stellar particles with a fixed time step of $\Delta t_{\rm SF} = 5 \, {\rm Myr}$\footnote{The time-step width in the simulation is much shorter than this, typically 1000 yr.}, while the simulation output times are not synchronized with this interval. The reassignment accounts for continuous star formation within each discrete time step.
%We therefore re-assign the stellar ages as the same as \citet{Nakazato:2023}. Specifically, for stellar particles younger than 15 Myr with age stamps at $T_1, T_2$, and $T_3$ ($T_1 < T_2 < T_3$), we randomly sample new ages within each interval. For example, for a star with age $T_1$, we randomly draw a new age within $[1, T_1]$ Myr and assign it as the updated age.

We note that we do not consider nebular continuum emission. Recent JWST observations suggest that nebular emission may dominate the UV continuum in some systems \citep[e.g.,][]{Cameron:2024}, and several theoretical studies have investigated the effect of nebular continuum \citep[e.g.,][]{Inoue:2011, Ceverino:2019, Katz:2025, Narayanan:2025_UVslope, Newman:2026}. In particular, \citet{Yanagisawa:2025} and \citet{Katz:2025} demonstrate that \HII\ regions around very young stars (ages $\sim$1--5 Myr) have dominant nebular continuum emission that makes UV slopes redder ($\beta_{\rm UV} \sim -2.5$), while stellar continuum alone produces bluer slopes ($\beta_{\rm UV} \sim -2.5$ to $-3$). However, the contribution of nebular continuum becomes negligible for stellar ages older than $\sim$10 Myr. Since most of the clumps we identified have ages $\gtrsim 30$ Myr (top panel of Figure \ref{fig:clump_properties}), we expect that nebular continuum does not significantly affect the UV slopes or the IR reemission calculated in our analysis\footnote{Clump ages are calculated as mass-weighted stellar ages. We have also calculated luminosity-weighted ages using the intrinsic UV luminosity at 1600 \AA\ as the weight, and confirmed that the difference from the mass-weighted ages is within $\simeq$ 3-5 Myr.}. We will investigate the contribution of nebular continuum for all FirstLight simulations in future work.

The SED (including both radiation field and dust emission) is sampled using 150 logarithmically spaced wavelength bins, covering the range from 0.1 to 1000\,\mum\footnote{For the instrument, the output SED is set to span 0.1-5000 \mum with 250 bins, noting that this range refers to the observed-frame wavelength (as opposed to the rest-frame wavelength used for the radiation field and dust emission grids).}. To ensure good convergence, a total of $10^7$ photon packets are emitted from the stars in each calculation. The numerical convergence with respect to the number of photon packets and the wavelength binning is verified in Appendix~\ref{sec:numerical_convergence}.

\subsubsection{Dust Spatial Distribution} \label{subsubsec:dust_spatial_distribution}
The spatial distribution of dust is imported from the hydrodynamical simulation by assuming that the dust distribution scales with that of metals, according to the fraction ${\rm DTM} = M_{\rm dust}/M_{\rm metal}$. We adopt a fixed value of ${\rm DTM} = 0.4$, which is used in several high-$z$ simulation studies \citep[e.g.,][]{Behrens:2018, Lovell:2021, Mushtaq:2023}. 
This value is also consistent with the analytical estimates by \citet{Dayal:2022}, who derived ${\rm DTM} \sim 0.32$-$0.4$ for observed $z \sim 7$ galaxies with dust continuum detections by ALMA (i.e., REBELS galaxies). Since the galaxies in our simulated sample have stellar masses of $M_\star \gtrsim 10^9\,M_\odot$ at $z \sim 7$, comparable to the ALMA-observed systems, the adopted DTM value is reasonable.

%Even though the value of ${\rm DTM}$ can affect RT outputs such as dust-to-gas mass ratio, dust temperatures, and IR luminosities, it is poorly constrained at high-$z$ observations \citep[e.g.,][]{Wiseman:2017} and models give a wide range of values \citep[e.g.,][]{Asano:2013, Aoyama:2017, Toyouchi:2025}. Adopting time-variable DTM is left for our future studies.

\subsubsection{Dust Absorption, Scattering and Reemission} \label{subsubsec:dust_graine_properties}
The size distribution and composition of the dust affect its absorption and scattering properties \citep[e.g.,][]{Draine:2014}. 
%We adopt dust \notice{types} of \citet{Weingartner_Draine:2001} that are appropriate for the extinction curve of Milky Way (MW) and Small Magellanic Cloud (SMC). 
We adopt two types of dust properties for the extinction curve of the Milky Way (MW) and Small Magellanic Cloud (SMC) \citep{Weingartner_Draine:2001}. 
The grain-size distribution of graphite, silicate, polycyclic aromatic hydrocarbons (PAH) for MW and SMC is sampled using 10/10/10 and 10/10/0 bins, respectively. The exact nature of dust grains in high-redshift galaxies is a matter of ongoing debate. Recent JWST observations have investigated dust attenuation curves in galaxies at $z > 4$ \citep{Fisher:2025, Ormerod:2025, Witstok:2023, Markov:2023, Markov:2025} and find that 20-25\% of REBELS and JADES samples exhibit evidence for a 2175\AA\, bump, which might be originated from carbonaceous grains including PAHs \citep[see ][]{Li:2024, Lin:2025, Nanni:2025} as seen in the MW extinction curve \citep{Cardelli:1989}, but not seen in the SMC extinction curve \citep{Gordon:2003}.  We, therefore, choose these two models and set the MW-dust model as a default model. The corresponding results for SMC-dust are shown in Appendix \ref{sec:SMC_dust}.

We consider the self-absorption of dust emission, which becomes important in dense clouds with large optical depths. In Appendix~\ref{sec:numerical_convergence}, we compare the results with and without self-absorption effects. We also include the transient heating function to calculate non-local thermodynamic equilibrium (NLTE) dust emission from stochastically heated small grains and PAH molecules \citep{Camps_Baes:2015}. SKIRT outputs the mass-weighted dust temperature\footnote{This temperature is referred to as "indicative temperature" in SKIRT terminology.} for each cell: 
\begin{equation}
T_{\rm cell} = \frac{\sum_i \rho_{{\rm cell}, i}T_{{\rm cell}, i}}{\sum_i \rho_{{\rm cell}, i}}, \label{eq:skirt_temperature}
\end{equation}
where $\rho_{{\rm cell}, i}$ is the mass density of dust type $i$, and $T_{{\rm cell}, i}$ is the dust temperature for type $i$ dust in a cell that is obtained through the energy balance equation. 
We also include dust heating from cosmic microwave background (CMB) radiation, whose temperature at high redshift increases as $T_{\rm CMB}(z) = (1+z)\times T_{{\rm CMB}(z=0)}$, and becomes nonnegligible \citep{da_Cunha:2015}. 

The radiative transfer calculations are iterated until the IR reemission luminosity converges within 3\%, or until a maximum of seven iterations is reached \footnote{The luminosity change at the 7th iteration has a median value of 5.97\% (16th--84th percentile: 3.94--8.67\%). We have confirmed that this level of non-convergence does not affect our results.}. An octree structure is used to construct the spatial grids of dust media, where cells are recursively subdivided until each contains less than $10^{-6}$ of the total dust mass. The highest grid level corresponds to a cell width of $\sim 4.9\, {\rm pc}$, approximately one-fourth of the minimal cell size in the hydrodynamical simulation. We have verified that this grid resolution is sufficient by checking that the true($=$input) and grid-discretized dust masses agree within 0.09\% and that the optical depth at rest-frame 5500 \AA\, agrees within 0.2\% along all three coordinate axes.

\begin{figure}
    \centering
    \includegraphics[width =\linewidth, clip]{\figdir/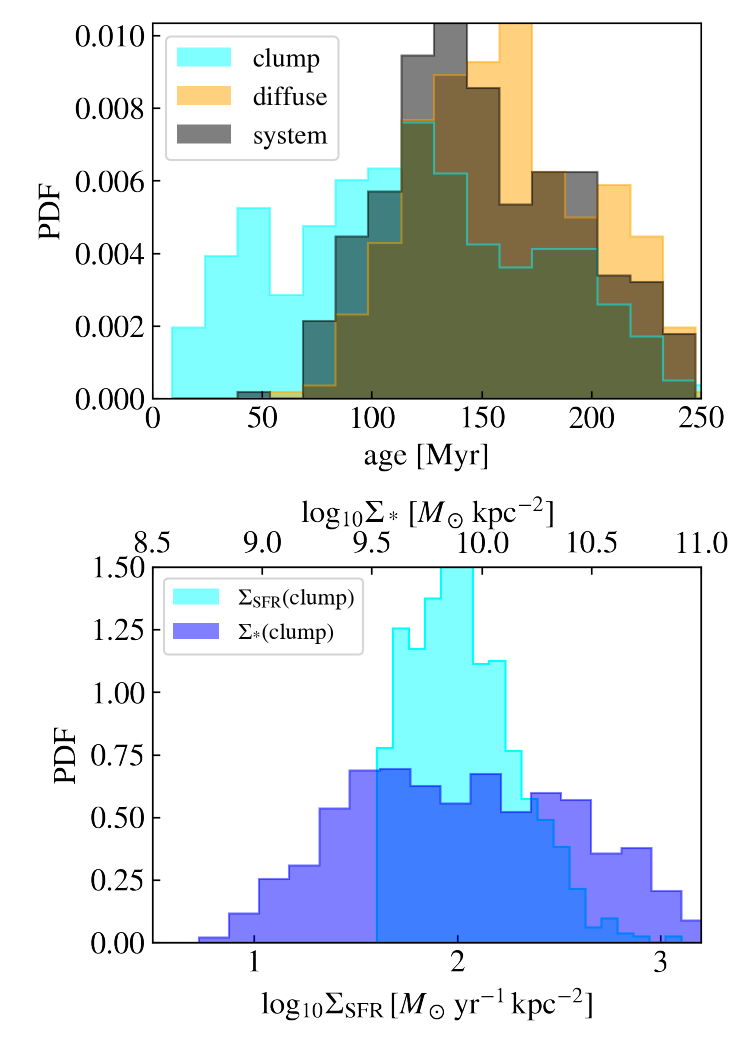}
    \caption{Histogram of clump properties. The top panel shows the mass-weighted stellar age distributions for each component: clumps (cyan), diffuse (orange), and system-integrated (black). The bottom panel shows the surface density distributions of SFR (cyan) and stellar mass (blue) for clump components. Note that these surface densities are only well-defined for clumps due to their compact nature.
}
    \label{fig:clump_properties}
\end{figure}

\begin{figure*}
    \centering
    \includegraphics[width =\linewidth, clip]{\figdir/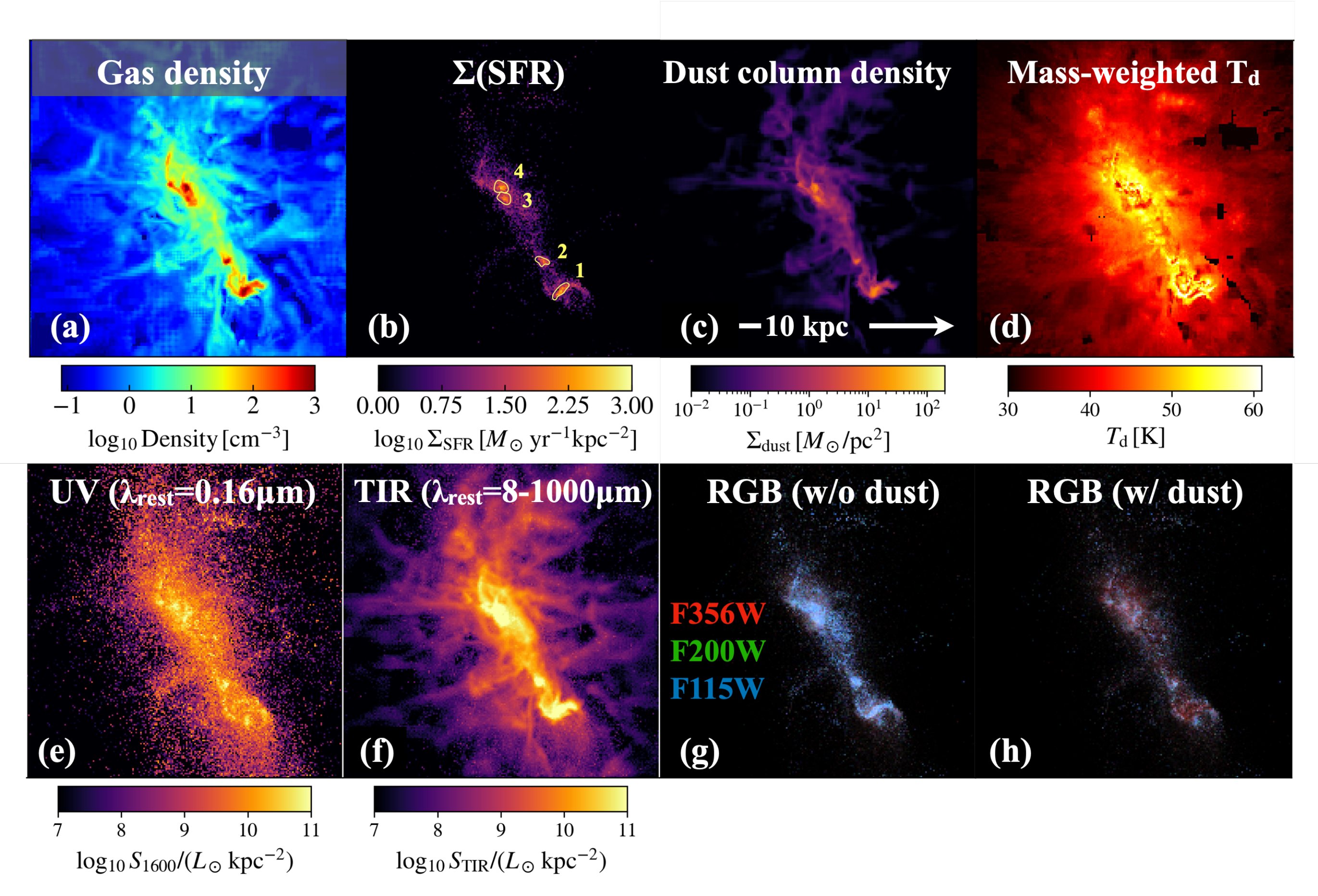}
    \caption{Projected distribution of simulated galaxy FL957 at $z=7.7$, one of the identified clumpy galaxies. {\it Upper panels:} (a) projected gas number density, (b) surface SFR density with yellow lines identifying star-forming clumps (see Section \ref{subsec:clump_identification}), (c) dust column density, and (d) mass-weighted dust temperature. {\it Lower panels:} (e) rest-frame UV surface brightness ($\nu L_\nu$ at 1600 \AA), (f) total IR emission integrated over $\lambda_{\rm rest} = 8$-$1000\,\mu{\rm m}$, (g) mock JWST/NIRCam three-color image (F115W, F200W, F356W) without dust attenuation, and (h) the same with dust attenuation. The surface brightness maps for each NIRCam band filter are computed by convolving the synthetic SEDs from SKIRT with the corresponding bandpass filters. Each panel shows a 10 kpc$\times$ 10 kpc region with a projected depth of 10 kpc.}
    \label{fig:projection_FL957_a0p115}
\end{figure*}

\subsection{Clump Identification} \label{subsec:clump_identification}
In order to quantitatively evaluate how dust properties vary spatially, we categorize each galaxy into three components: (1) the \textit{system-integrated} component, representing the entire galaxy within a projected $10\,{\rm kpc} \times 10\,{\rm kpc}$ region; (2) individual \textit{star-forming clumps}, identified based on SFR surface density thresholds; and (3) the \textit{diffuse} component, defined as the system-integrated region with identified clumps masked out, which includes tidal tails, bridges, and inter-clump regions. This categorization allows us to investigate whether and how local physical conditions (dust-star geometry, dust column density) affect dust attenuation and reemission properties such as temperature, UV slope, and attenuation curve shape. We note that this clump identification is based on intrinsic SFR surface density maps, which does not always correspond to observational clump identification methods affected by point spread function (PSF), noise, and bandpass filters. We discuss this caveat in detail in Section \ref{subsec:caveats}.

We identify star-forming clumps based on the surface density of the SFR, following \citet{Nakazato:2024}. Here, we briefly summarize the procedure. Clumps are identified in two-dimensional “images” constructed by projecting a cubic region with a fixed side length of 10~kpc along the line of sight. The gas and stellar masses are re-assigned to a uniform grid with a cell size of $\Delta = 50\,{\rm pc}$.

To identify $\sim$100~pc-scale star-forming clumps comparable to those observed with JWST, we adopt a threshold SFR surface density of $\Sigma_{\rm SFR} = 10^{1.5}\,M_\odot\,{\rm yr^{-1}\,kpc^{-2}}$ per grid. Clumps are identified above this threshold using a dendrogram technique \citep{Rosolowsky:2008}. Groups with at least $N_{\rm grid} = 16$ grid cells are identified as clumps, allowing us to detect clumps with radii of $\gtrsim 100$~pc and ${\rm SFR} \gtrsim 1\,M_\odot\,{\rm yr^{-1}}$, corresponding to observed clumps with intrinsic \OIII~5007 fluxes of $\sim 3\times 10^{-18}\,{\rm erg\,s^{-1}\,cm^{-2}}$\footnote{This flux value is derived from the \OIII 5007\AA-SFR relation shown in \citet{Nakazato:2024} (see their Eqs. (6) and (7)). The flux $\sim 3\times 10^{-18}\, {\rm erg\,s^{-1}\, cm^2}$ yields a signal-to-noise ratio $>$ 5 for a typical exposure time of $\sim 10^4$ s adopted in several JWST observations \citep[e.g.,][]{Hashimoto:2023, Arribas:2024, Jones:2026_B14-65666, Matthee:2023}. A more detailed description is given in \citet{Nakazato:2024}.}, one of the brightest emission lines in high-$z$ galaxies. We note that in our clump identification, we do not distinguish between merging companions and star-forming clumps formed in tidal tails during peri-center passages. This approach is consistent with observational methods, where clumpy systems are identified without distinguishing between these two populations. 

%In our previous work \citep{Nakazato:2024}, we also examined the virial parameters of the identified clumps, defined as $\alpha_{\rm vir} \equiv 5\sigma_{\rm clump}^2 R_{\rm clump}/(G M)$, where $\sigma_{\rm clump}$ and $M$ are the velocity dispersion and baryonic mass of each clump, respectively. The typical value is $\alpha_{\rm vir} \sim 0.2$, indicating that the clumps are gravitationally bound.

%Among 4092 snapshots at $z=9.5$--$5.5$, we identify 376 snapshots in the clumpy phase, defined as systems containing at least two clumps. 
Among 4092 snapshots at $z=9.5$--$5.5$, we identify 376 snapshots that contain at least two clumps. In total, 1059 clumps are identified in these systems. Figure~\ref{fig:clump_properties} shows the physical properties of clumps in these systems. The top panel shows the mass-weighted stellar age distributions for each component: clumps (cyan), diffuse regions (orange), and system-integrated (black), with mean values of 118~Myr, 158~Myr, and 145~Myr, respectively. The clump age distribution is bimodal, with peaks at $\lesssim 50$~Myr and $\gtrsim 50$~Myr; the younger population primarily forms in tidal tails during mergers, as explained in \citet{Nakazato:2024}. The bottom panel shows the stellar mass surface density (blue, upper x-axis) and SFR surface density (cyan, lower x-axis) distributions for clumps. The median values are $\log_{10}\Sigma_{*} \,[M_\odot\,{\rm kpc^{-2}}]= 9.9\,$ and $\log_{10}\Sigma_{\rm SFR} \, [M_\odot\,{\rm yr^{-1}}\,{\rm kpc^{-2}}]= 2.0$, respectively, indicating compact star-forming clumps. These 376 clumpy systems are used for the statistical analysis in the following sections\footnote{We note that these 376 clumpy systems are not statistically independent, as the same 62 halos are traced across multiple snapshots. However, since the 376 clumpy systems are drawn from 62 halos spanning a wide range of evolutionary stages and stellar masses with 1.84$\times 10^9 - 1.54\times 10^{11} \, M_\odot$, we expect the statistical trends reported in this work to be robust. A detailed quantitative assessment of potential statistical biases is beyond the scope of this study.}.

\begin{figure*}
    \centering
    \includegraphics[width =\linewidth, clip]{\figdir/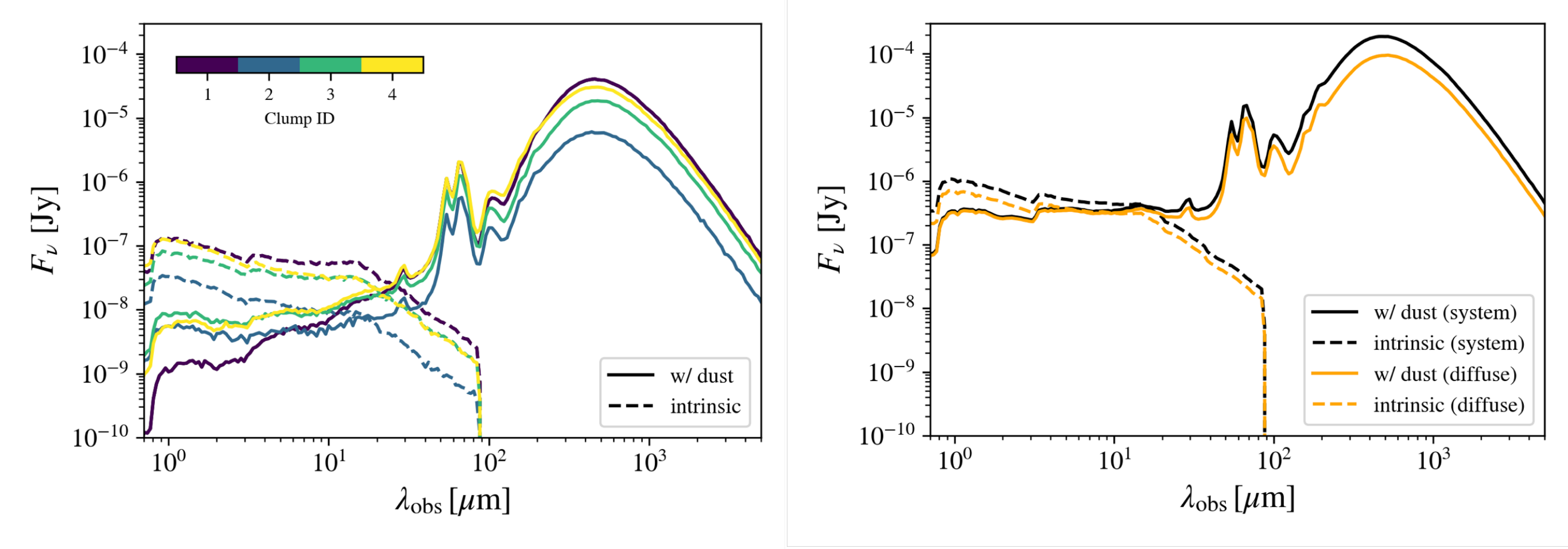}
    \caption{{\it Left:} SEDs for the four clumps. The dashed and solid lines indicate the intrinsic stellar continuum and the attenuated (with reemission) SEDs, respectively. {\it Right:} SEDs for the system-integrated (black) and diffuse (orange) components. The system-integrated SED is calculated over the entire 10 kpc $\times$10 kpc region, while the diffuse component SED is obtained by subtracting the clump emission from the system-integrated value.
}
    \label{fig:SED_FL957_a0p115}
\end{figure*}
\section{Result} \label{sec:result}
\begin{figure*}
    \centering
    \includegraphics[width =\linewidth, clip]{\figdir/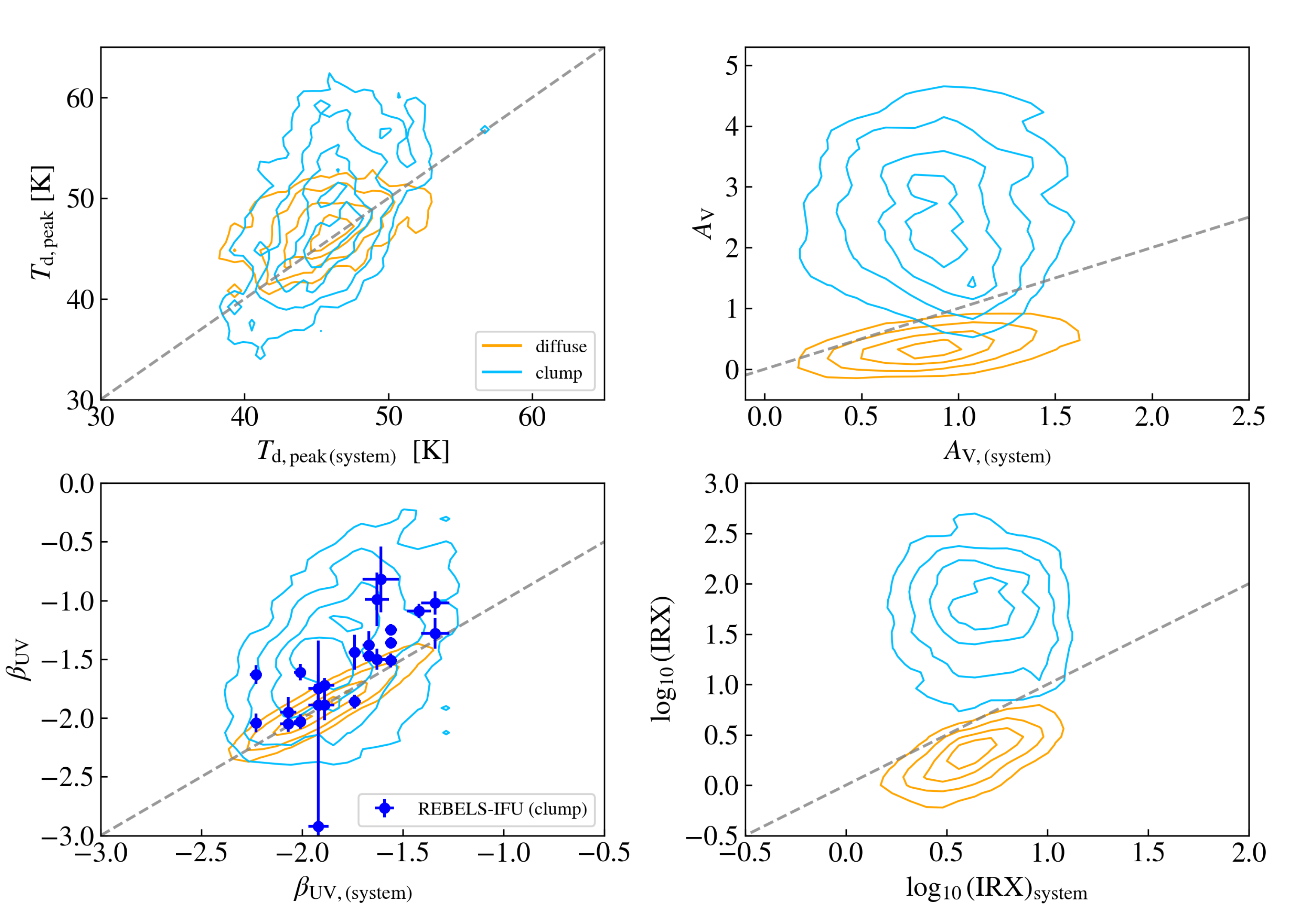}
    \vspace{-4mm}
    \caption{Comparison of dust observables between system-integrated values and individual components for all identified clumpy galaxies at $z=6-9$. The x-axis of each panel shows the system-integrated value calculated over 10 kpc $\times$ 10 kpc regions, while the y-axis shows values for individual clumps (cyan) and diffuse components (orange). Contours enclose 20\%, 50\%, 80\%, and 95\% of the data. {\it Top left:} peak dust temperature ($T_{\rm d, peak}$). {\it Top right:} V-band attenuation ($A_{\rm V}$). {\it Bottom left:} UV slope ($\beta_{\rm UV}$). For comparison, we also plot 
observational results from the REBELS-IFU survey \citep{Fisher:2025, Komarova:2026}, comparing system-integrated and clump components. {\it Bottom right:} infrared excess (IRX $\equiv L_{\rm IR}/L_{\rm UV}$). The gray dashed line in each panel indicates the one-to-one relation.
}
    \label{fig:dust_properties_statistical}
\end{figure*}
\subsection{Dust Properties of an Individual Clumpy Galaxy: FL957 at $z=7.7$} \label{subsec:case_study} 
 In this section, we present one representative example: FL957 at $z=7.7$, which was also introduced in \citet{Nakazato:2024} and has four clumps within a 10 $\times$ 10 kpc region. 
The physical properties of each clump, such as stellar mass, mass-weighted age, and clump radius, are tabulated in Table \ref{table:FL957_physical_properties}. 
The clumpy systems were formed during the merger phase. 

Figure \ref{fig:projection_FL957_a0p115} shows the projected maps of physical properties of FL957. 
%Each panel shows the distributions of (a) gas number density, (b) surface star formation rate density, (c) dust column density, (d) mass-weighted dust temperature, (e) rest-frame UV surface brightness at 1600 \AA, (f) total IR surface brightness (integrated over 8--1000 $\mu$m), (g) RGB composite without dust attenuation for NIRCam filters (F115W, F200W, F356W), and (h) RGB composite with dust attenuation. 
Among the four clumps, two clumps formed within a tidal tail produced by the peri-center passage of two galaxies. Each clump has high gas densities of $\simeq 200$--$400\, {\rm cm^{-3}}$ as shown in panel (a), triggering intense star formation with surface densities $\simeq 60$--$115\, M_\odot\, {\rm yr^{-1}\, kpc^{-2}}$ (panel (b)). Such clumps are bright in \OIII~5007\AA\, (see Figure 2 of \citet{Nakazato:2024}). A fraction of these newly formed stars end their lives as Type II SNe, enriching the surrounding gas with metals and producing large amounts of dust. The distribution of dust column density in panel (c) therefore traces the distributions of gas density and surface SFR density, reflecting the enhanced dust production in regions of bursty star formation. 
%The mass-weighted dust temperature in panel (d) shows 
Panel (d) shows the mass-weighted dust temperature, where the temperature of each cell (Eq. (\ref{eq:skirt_temperature})) is averaged along the line of sight using dust mass as weights. We see elevated temperatures ($T_{\rm d, mass} \sim 50$--60 K) in the clumps, where intense star formation heats the dust locally. The surrounding diffuse regions exhibit lower temperatures ($T_{\rm d, mass} \sim 40$--45 K), reflecting the lower radiation field intensities away from the star-forming sites. The dust in these regions absorbs UV/optical photons and reemits them as IR radiation, and the IR map in panel (f) also corresponds to this distribution. The UV emission in panel (e) is a noisier image in comparison to the IR image, due to the scattering effect of UV photons. In the NIRCam RGB images (panels (g) and (h)), although the intrinsically UV-bright regions are colocated with the {\it physical} star-forming clumps, the dust-attenuated images show smaller clumps. This is because the clumps have high dust column densities ($A_{\rm V} = 1.1$--$2.8$) that heavily attenuate UV radiation, whereas the surrounding diffuse gas has much lower attenuation ($A_{\rm V} \sim 0.09$).
Such effects are seen in several simulated galaxies \citep{Ceverino:2026}, but also in observations \citep[e.g.,][Hagimoto et al. in prep.]{Bowler:2022, Rowland:2024, Crespo-Gomez:2024}.

Figure \ref{fig:SED_FL957_a0p115} shows the SEDs for the clumpy galaxy FL957 at $z=7.7$. The left panel shows the SEDs for the four clumps, and the right panel shows the SEDs for the system-integrated and diffuse components. The dashed and solid lines represent intrinsic SEDs and dust-attenuated (with reemission) SEDs, respectively. We note that although the diffuse component is defined as all pixels excluding the identified clump pixels (Section \ref{subsec:clump_identification}), the inclusion of faint pixels does not significantly affect the total SED, as the top 50\% brightest pixels in the diffuse component account for 87\% and 97\% of the total $L_{\rm UV}$ and 
$L_{\rm IR}$, respectively.

From the SEDs, we calculate several dust observables as summarized in Table \ref{table:FL957_dust_properties}: UV slope ($\beta_{\rm UV}$), $L_{\rm IR}$, $L_{\rm UV}$, IRX, $A_{\rm V}$ ($A_{\rm UV}$), and $T_{\rm peak}$. The UV continuum slope $\beta_{\rm UV}$ is widely used as an indicator of UV continuum attenuation. Thanks to JWST’s unprecedented sensitivity in infrared imaging and spectroscopy extending up to $\lambda_{\rm obs} \sim 5\, \mu{\rm m}$, robust measurements of the UV slope at $z\gtrsim 7$ have recently become available for the first time \citep[e.g.,][]{Topping:2022, Cullen:2023,  Morales:2024, Cullen:2024, Topping:2024, Dottorini:2025, Saxena:2026}. We calculate $\beta_{\rm UV}$ in the same way as in \citet{Liang:2021_IRX_beta}: 
\begin{equation}
\beta_{\rm UV} = \frac{\log_{10}(f_{\lambda, {\rm 1230\mathrm{\mathring{A}}}})-\log_{10}(f_{\lambda, {\rm 3200\mathring{A}}}) }{\log_{10}(\lambda_{\rm 1230\mathring{A}})- \log_{10}(\lambda_{\rm 3200\mathring{A}})}, \label{eq:UV_slope_Liang21}
\end{equation}
where $f_{\lambda, {\rm 1230\mathring{A}}}$ and $f_{\lambda, {\rm 3200\mathring{A}}}$ are the flux densities at rest-frame 1230 \AA\, and 3200 \AA, respectively. We adopt this wavelength range to avoid the contamination of the UV bump effect at $\sim 2175$ \AA, which systematically leads to underestimations of the UV slope \citep{Inoue:2006, Narayanan:2018_IRX_beta}. 

For the dust temperature, we calculate the peak dust temperature $T_{\rm peak}$, which is obtained from the Wien displacement law from the rest-frame wavelength at which the infrared flux density ($F_\nu$) peaks ($\lambda_{\rm peak}$)\footnote{The peak flux density is taken as $B_\nu$, not $B_\lambda$.}. This is defined as 
\begin{equation}
T_{\rm d, peak} = \frac{2.898 \times 10^3\, \mu{\rm m}\, {\rm K}}{\lambda_{\rm peak}}, \label{eq:peak_temperature}
\end{equation}
which follows the relation for a graybody with a dust emissivity index $\beta_{\rm d}=2$, i.e., $L_\nu \propto (1-e^{-\tau_\nu}) B_{\nu} \sim \nu^{\beta_{\rm d}}B_\nu = \displaystyle \frac{\nu^{3+\beta_{\rm d}}}{e^{h_{\rm p}\nu/(k_{\rm B}T)}-1}$. This measure has been used in many observational studies \citep[e.g.,][]{Casey:2018, Schreiber:2018, Burnham:2021, Witstok:2022, Mitsuhashi:2024}. 
%A more realistic form is the two-component dust SED model, consisting of a modified blackbody (MBB) function for cold dust and a power-law component for warmer dust \citep{Casey:2012}. Nevertheless, an optically thin MBB function at local equilibrium temperature is still a good approximation for the local dust emissivity at rest-frame $\lambda_{\rm rest} > 30\, \mu{\rm m}$ where non-LTE effects are negligible. 
Note that $T_{\rm d, peak}$ is simply a proxy for $\lambda_{\rm peak}$, and recent studies suggest that such a single dust temperature does not always recover the true values of dust mass and IR luminosity \citep[e.g.,][]{Sommovigo:2025_Td}. However, we adopt this definition for direct comparison with other theoretical and observational studies.

We define IRX as 
\begin{equation}
{\rm IRX} \equiv \frac{L_{\rm IR}}{L_{\rm UV}} = \frac{\int^{1000\, \mu\rm m}_{8\, \mu \rm m} L_\lambda\, {\rm d}\lambda}{\lambda_{\rm 1600\AA}L_{\lambda, 1600\AA}}, 
\end{equation}
where all wavelengths are in the rest frame. 
%These dust properties of each component for FL957 at $z=7.7$ are tabulated in Table \ref{table:FL957_dust_properties}.

We find that the intrinsic UV slope $\beta_{\rm UV, 0}$ is almost the same for all components with $\beta_{\rm UV, 0} = -2.8$ to $-2.7$, but the attenuated $\beta_{\rm UV}$ is very different among components. In particular, Clumps 1 and 4 with large gas densities of $\sim 400$--$500\, {\rm cm^{-3}}$ have large differences with $\Delta \beta = \beta_{\rm UV} - \beta_{\rm UV, 0} = 0.7$, $1.0$, and the corresponding UV attenuation is also large with $A_{\rm UV} \sim 2.95$--$4.60$. In contrast, the diffuse component has a very low gas density with a median value of $\langle n_{\rm gas} \rangle = 1.9\, {\rm cm^{-3}}$\footnote{The median value is calculated from the 2D pixels of the projection map in Figure \ref{fig:projection_FL957_a0p115}.}, and the attenuated slope is $\beta_{\rm UV} = -2.35$ ($\Delta \beta = 0.35$) with $A_{\rm UV} = 0.64$. This result is reflected in the system-integrated value, with a relatively blue slope of $\beta_{\rm UV} = -2.34$. For the V-band attenuation $A_{\rm V}$, clumps have large values ($A_{\rm V} = 1.12$--$2.84$), while the diffuse and system components have $A_{\rm V} = 0.09$ and $0.38$, respectively. We further discuss these values in terms of attenuation curves in Section \ref{subsec:attenuation_curves}. 

For the dust peak temperature, we find that the four clumps in FL957 have temperatures 4-10 K higher than the diffuse component. This is because clumps have radiation sources in compact regions, and the dust temperature scales as $T_{\rm d} \propto u_*^{1/(4+\beta_{\rm d})} \sim (L_*/r^{2})^{1/(4+\beta_{\rm d})}$, where $u_*$, $L_*$, and $r$ are the radiation energy density, luminosity of sources, and the distance between the radiation source and dust grains, respectively. Spatially resolved dust temperatures for $z\gtrsim 5$ galaxies (and quasars) have recently been investigated by high-resolution ALMA observations \citep[e.g.,][]{Tsukui:2023, Meyer:2025, Villanueva:2024, Fernandez-Aranda:2025}. In particular, the ALMA-CRISTAL survey found that the bridge regions between two clumps have 5 K lower dust temperatures than the clumps, also showing similar trends.

Finally, regarding IRX, clumps have large values of $\log {\rm IRX} = 0.68$--$1.99$, comparable to local (U)LIRGs \citep[e.g.,][]{Goldader:2002, Howell:2010} and IR-selected dusty star-forming galaxies (DSFG) at $z\sim2$ \citep{Casey:2014_IRX_beta}. They are up to around 2 dex larger than the diffuse and system components with $\log {\rm IRX} = 0.07$ and $0.34$, respectively. We will further discuss this in terms of the IRX--$\Delta \beta$ relation in Section \ref{subsec:toy_model}. 

\begin{table*}[ht]
\centering
\renewcommand{\arraystretch}{1.2}
\begin{tabular}{lcccccc}
\hline
Component & Number & $\langle\beta_{\rm UV}\rangle$ & $\langle\beta_{\rm UV,0}\rangle$ & $\langle \log_{10}({\rm IRX})\rangle$ 
& $\langle L_{\rm UV,1600} \rangle$ [$L_\odot$] & $\langle L_{\rm IR} \rangle$ [$L_\odot$] \\
\hline \hline
Clump   & 1059 & $-1.48_{-0.42}^{+0.48}$ & $-2.49_{-0.21}^{+0.29}$ & $1.75_{-0.41}^{+0.35}$ & $1.01_{-0.61}^{+2.10}\times 10^{9}$ & $5.02_{-3.17}^{+18.9}\times 10^{10}$ \\
Diffuse & 376  & $-1.90_{-0.20}^{+0.20}$ & $-2.36_{-0.16}^{+0.16}$ & $0.30_{-0.22}^{+0.20}$ & $1.10_{-0.38}^{+0.77}\times 10^{11}$ & $2.12_{-1.06}^{+2.61}\times10^{11}$ \\
System  & 376  & $-1.89_{-0.20}^{+0.20}$ & $-2.36_{-0.17}^{+0.16}$ & $0.62_{-0.16}^{+0.19}$ & $1.13_{-0.39}^{+0.83}\times 10^{11}$ & $4.86_{-2.29}^{+4.56}\times 10^{11}$ \\
\hline 
\noalign{\vspace{1mm}}
\hline
Component & $\langle T_{\rm d, peak} \rangle$ [K] & $\langle A_{\rm UV}\rangle$ & $\langle A_V \rangle$ & $\langle S\rangle$ \\
\hline \hline
Clump   & $47.68_{-4.9}^{+6.0}$ & $4.13_{-0.98}^{+0.85}$ & $2.50_{-0.90}^{+0.96}$ & $1.60_{-0.21}^{+0.45}$ \\
Diffuse & $47.09_{-2.2}^{+2.1}$ & $1.00_{-0.28}^{+0.26}$ & $0.34_{-0.16}^{+0.21}$ & $2.90_{-0.71}^{+1.30}$ \\
System  & $45.50_{-2.5}^{+2.4}$ & $1.59_{-0.31}^{+0.34}$ & $0.87_{-0.28}^{+0.27}$ & $1.84_{-0.26}^{+0.36}$ \\
\hline
\end{tabular}
\caption{Summary of median dust observables for clumpy galaxies identified in FirstLight simulations with a $40\, {\rm cMpc}/h$ box at $z=6$--9.5. We identified 376 clumpy systems containing 1059 clumps in total. Here, $\beta_{\rm UV}$ ($\beta_{\rm UV, 0}$) is the UV slope for the attenuated (intrinsic) SED. 
Dust temperature is calculated as the peak dust temperature $T_{\rm d, peak}$ 
(see Equation \ref{eq:peak_temperature}). The attenuation curve slope is 
defined as $S \equiv A_{\rm UV}/A_{\rm V}$. IRX is the infrared excess, 
defined as ${\rm IRX} \equiv L_{\rm IR}/L_{\rm UV}$. Each value shows the 
median for each component. We categorize three components: clumps, diffuse regions, and system-integrated values. The superscript and subscript values indicate the 16th-84th percentile range.} \label{table:statistical_dust_properties}
\end{table*}
\subsection{Dust observables: clumps vs system-integrated values} \label{subsec:statistical_dust_properties}
In the previous section, we have focused on the dust observables of individual clumps in FL957. In this section, we examine the statistical dust observables using all 376 clumpy galaxies identified in our simulations. For each system, we show the results of the z-axis projection, as star-forming clumps are identified along this axis (see Section \ref{subsec:clump_identification}). Since the 376 clumpy systems have diverse morphologies, and merger-induced clumps at high-redshifts are distributed in 3D rather than confined to a single plane, the fixed z-axis projection is effectively equivalent to a random orientation.
%Figure \ref{fig:dust_properties_statistical} compares dust properties between system-integrated values and individual components (clumps and diffuse regions). 
Figure \ref{fig:dust_properties_statistical} shows various dust observables of the clump and diffuse components, plotted as a function of the corresponding system-integrated values.
We compare four key quantities: dust peak temperature, V-band attenuation, UV slope, and IRX.

The top-left panel shows that clumps exhibit larger scatter in dust temperature 
than the diffuse components and systematically show higher dust temperatures. 
Some clumps have $T_{\rm d, peak} \sim 65$ K, approximately 20 K higher than typical system-integrated values ($T_{\rm d, peak} \sim 45$ K), while the median values for clump and diffuse components are almost the same, $\sim 45$--$47$ K (see Table \ref{table:statistical_dust_properties}). 
These clumps with elevated dust temperatures reflect compact star formation regions, as discussed in Section \ref{subsec:case_study}. Interestingly, 23\% of clumps have lower $T_{\rm d, peak}$ than the system-integrated values. These clumps have higher SFR surface densities (median: $\Sigma_{\rm SFR}= 172 \, M_\odot\, {\rm yr^{-1}\, kpc^{-2}}$) than the remaining clumps (median: $\Sigma_{\rm SFR}= 90 \, M_\odot\, {\rm yr^{-1}\, kpc^{-2}}$), leading higher dust column densities and stronger dust-shielding, which results in lower dust temperatures in the clumps.

For $A_{\rm V}$ in the top-right panel, clumps show large scatter and typically exhibit higher attenuation, with values up to 4 mag larger than the corresponding system-integrated values. In contrast, diffuse components lie systematically below the one-to-one line, with attenuation up to 0.7 mag lower than the system values. UV slopes in the bottom-left panel show similar trends: clumps have redder slopes with $\beta_{\rm UV, clump}-\beta_{\rm UV, sys}$ up to $\sim 1$, while diffuse components have slopes similar to the system-integrated values. Recent JWST observations have begun to provide spatially resolved $\beta_{\rm UV}$ measurements using NIRCam photometry \citep[e.g.,][]{Sugahara:2025} and NIRSpec IFU spectroscopy \citep[e.g.,][]{Marconcini:2024, Komarova:2026}. For example, \citet{Fisher:2025} and \citet{Komarova:2026} reported system-integrated and clump-resolved $\beta_{\rm UV}$ values for 12 REBELS-IFU sample galaxies. From their measurements, we calculate 
clump-system differences of $\beta_{\rm UV, clump}-\beta_{\rm UV, sys} 
\simeq 0.2$ (up to 0.6; blue plots in the bottom-left panel of 
Figure \ref{fig:dust_properties_statistical}), consistent with our 
simulation results. 

For IRX in the bottom-right panel, clumps also exhibit large scatter and systematically show larger IRX values than the system-integrated values, with offsets up to $\sim 2$ dex. In contrast, diffuse components always have smaller IRX than the system values, with offsets up to $\sim 0.5$ dex. These distinct IRX distributions will be discussed in terms of the IRX--$\Delta \beta$ relation in Section \ref{subsec:toy_model}. The median values of these dust observables for each component are summarized in Table \ref{table:statistical_dust_properties}.

The large scatter and systematic offsets in dust observables between different components suggest that commonly used system-integrated measurements may not adequately represent the dust observables of individual star-forming regions. These component-level variations have important implications for interpreting spatially resolved observations and for understanding the physical mechanisms governing dust attenuation and emission, which we explore in the following sections.

\begin{figure}
    \centering
    \includegraphics[width =\linewidth, clip]{\figdir/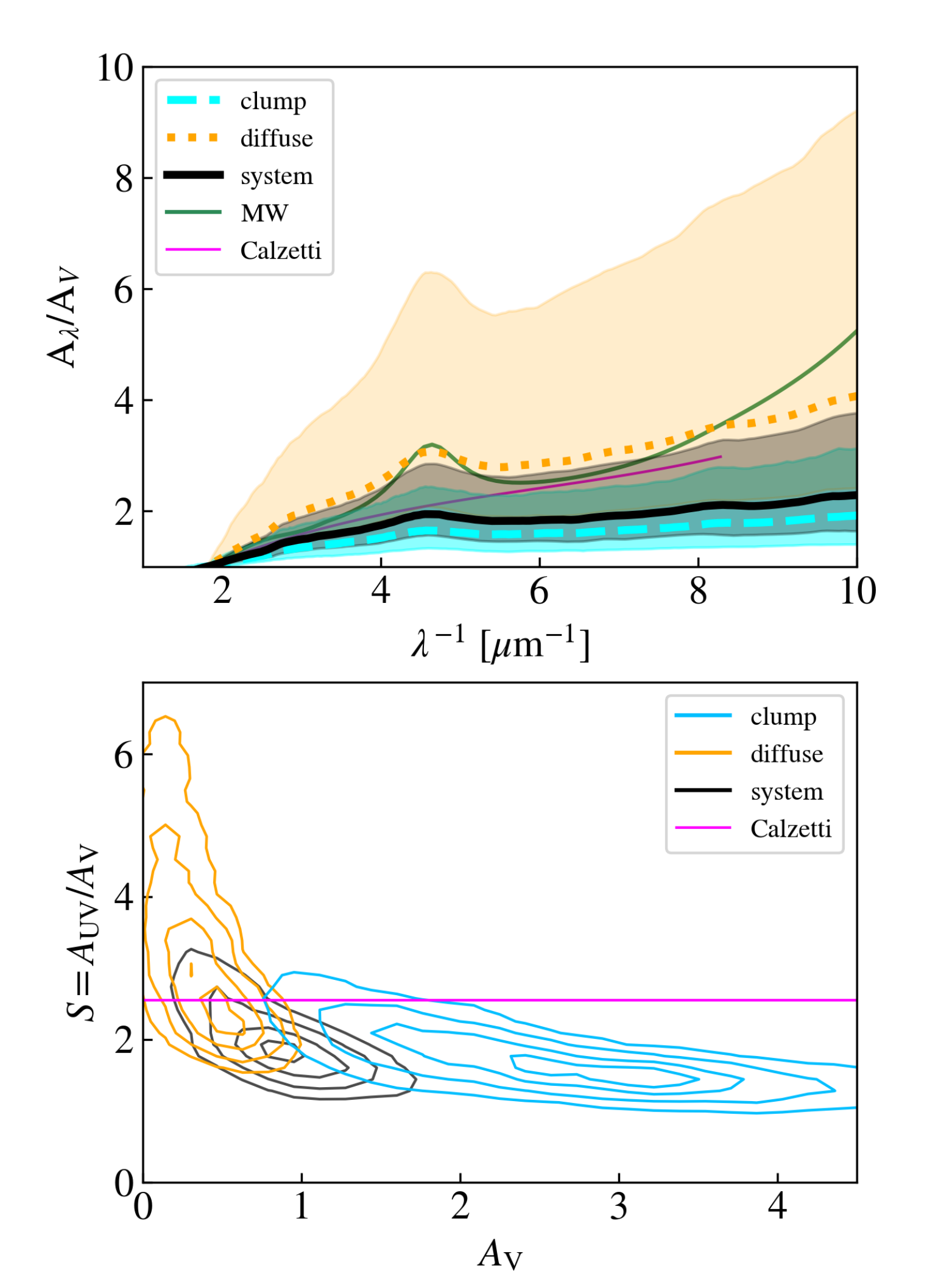}
    \vspace{-6mm}
    \caption{{\it Top panel:} Dust attenuation curves for the clumpy galaxies we identified in our simulations. The black solid line shows the median system-integrated attenuation curve, colored dashed lines show the median attenuation curves of individual clumps, and the orange dotted line shows the median of the diffuse component. Each shaded region shows the 5-95\% range. For comparison, we plot the Calzetti attenuation curve \citep[magenta, ][]{Calzetti:2000} and MW extinction curve \citep[green, ][]{Cardelli:1989}.
{\it Bottom panel:} Attenuation curve slope ($S \equiv A_{\rm UV}/A_{\rm V}$) versus V-band dust attenuation ($A_{\rm V}$) for all identified clumpy galaxies at $z=6-9$. The black, cyan, and orange contours represent the system-integrated, individual clumps, and diffuse components, respectively. Contours enclose 20\%, 50\%, 80\%, and 95\% of the data. The horizontal magenta line indicates the Calzetti attenuation law.
}
    \label{fig:attenuation_curve}
\end{figure}

\subsection{Attenuation curves} \label{subsec:attenuation_curves}
We investigate the dust attenuation curves for different components. The top panel of Figure \ref{fig:attenuation_curve} shows the attenuation curves for all identified clumpy galaxies, including individual clumps, diffuse components, and system-integrated values. The attenuation curves are derived from the SEDs, as exemplified for FL957 in Figure \ref{fig:SED_FL957_a0p115}. The bottom panel shows the attenuation curve slope ($S \equiv A_{\rm UV}/A_{\rm V}$) versus V-band dust attenuation ($A_{\rm V}$) for all components.

For the system-integrated attenuation curves (solid black lines), we find that they are grayer than the Calzetti attenuation curve \citep[$S=2.55$, ][]{Calzetti:2000}, with a median slope of $\langle S \rangle = 1.84$. This trend toward grayer curves is consistent with recent JWST observations of $z \gtrsim 6$ galaxies \citep[e.g.,][]{Markov:2025, Shivaei:2025}. 

When we examine spatially resolved attenuation curves, however, we find significant variations among components. Clumps exhibit grayer (flatter) attenuation curves than the system-integrated curves, while diffuse components show much steeper curves. From the $S$--$A_{\rm V}$ relation (bottom panel), clumps systematically occupy the region with lower slopes (grayer curves) and higher $A_{\rm V}$ values, while diffuse components show higher slopes (steeper curves) with lower $A_{\rm V}$ values. The median slopes for clumps, diffuse components, and system-integrated values are $\langle S \rangle = 1.60$, $2.90$, and $1.84$, respectively (see also Table \ref{table:statistical_dust_properties}). 

Similar trends of grayer curves in star-forming regions compared to system-integrated values have been observed in local galaxies \citep{Chevallard:2013, Decleir:2019}, and have also been found in simulations of local isolated galaxies \citep{Matsumoto:2026}. 
%However, these component-wise differences in EoR galaxies are quantitatively derived 
However, a systematic and quantitative component-by-component analysis of these differences in EoR galaxies is presented here for the first time. The physical origin of these systematic differences is discussed in Section \ref{subsec:discussion_attenuation}.

\section{Discussion} \label{sec:discussion}
\subsection{Physical Origin of Component-wise Attenuation Curves} 
\label{subsec:discussion_attenuation}
We have shown that our $z = 6$--$9$ galaxies exhibit grayer system-integrated attenuation curves than the Calzetti curve, with significant component-wise variations: clumps show the grayest curves ($\langle S \rangle = 1.60$), while diffuse components show the steepest ($\langle S \rangle = 2.90$). We now investigate the physical origin of these systematic differences.

\citet{Markov:2025} argue that grayer attenuation curves at high redshift compared to local galaxies can be reproduced by three factors: (i) larger $A_{\rm V}$ attenuation, (ii) larger dust grains, and (iii) compact dust-star geometry. 
For scenario (i), in regions with large dust attenuation (large $A_{\rm V}$), 
dust absorption dominates over scattering processes, resulting in grayer 
attenuation curves as shown in Appendix \ref{sec:scattering_effect} \citep[also, ][]{Narayanan:2018_attenuation_curve, 
Lin:2021, Matsumoto:2026}. However, the $S$--$A_{\rm V}$ relation 
(Figure \ref{fig:attenuation_curve}, bottom panel) shows that over 80\% 
of our system-integrated curves are grayer than low-z ($z \sim 1$--$3$) 
star-forming galaxies at the same $A_{\rm V}$ \citep[Figure 8 of ][]{Shivaei:2025}. This 
indicates that scenario (i) is unlikely to be the dominant explanation.
In scenario (ii), larger dust 
grains produce grayer extinction curves, leading to grayer attenuation curves. 
However, we adopt fixed dust compositions for MW-like dust 
\citep{Weingartner_Draine:2001} (also SMC-like dust in 
Appendix \ref{sec:SMC_dust}), and scenario (ii) does not apply to our study. 
Therefore, scenario (iii), compact dust-star geometry, is the most favored 
explanation for the grayer system-integrated curves. In our clumpy 
(i.e., extended) galaxies at $z = 6$--$9$, star-forming regions are compact 
and embedded in dense dusty environments, resulting in gray attenuation curves 
even at moderate $A_{\rm V}$ values.

The component-wise differences can be understood as a combination of scenarios (i) and (iii). Clumps exhibit grayer attenuation curves due to their higher optical depths, where dust absorption dominates over scattering, and their compact dust-star geometry. In these high optical depth environments, scattered photons are more likely to be absorbed before escaping, suppressing scattering effects and resulting in grayer curves \citep{Narayanan:2018_attenuation_curve}. In contrast, diffuse components have low optical depths, making scattering effects more pronounced. In optically thin regions, scattering can effectively reduce $A_{\rm V}$ while maintaining higher $A_{\rm UV}$, resulting in steeper slopes ($A_{\rm UV}/A_{\rm V}$). The extended geometry of diffuse components also contributes to steeper curves.

In summary, clumps exhibit grayer attenuation curves due to their higher optical depths (i) and compact geometry (iii), where scattering effects are suppressed. Diffuse components, with their lower optical depths and extended geometries, show steeper curves dominated by scattering. In the next section, we use the IRX--$\Delta \beta$ relation to further 
distinguish qualitatively between the effects of optical depth and dust-star geometry. These systematic differences have important implications for interpreting spatially resolved dust observations and for accurate SED modeling of high-redshift clumpy galaxies.

\subsection{Quantitative Evaluation of Dust-Star Geometry Using IRX-$\Delta \beta$} \label{subsec:toy_model}
As shown by \citet{Seon_Draine:2016} and \citet{Lin:2021}, quantitatively 
evaluating the effects of both dust optical depth and dust-star geometry 
from attenuation curves alone is challenging. Because attenuation curves 
involve ratios of attenuations at different wavelengths, they are affected 
by the wavelength dependence of scattering. Properly accounting for photons 
scattered into the line of sight from outside requires full 3D radiative 
transfer calculations. However, the IRX--$\Delta \beta$ relation, where 
$\Delta \beta$ denotes the difference between the attenuated and intrinsic 
UV slopes, provides a more direct diagnostic of geometry and optical depth. 
When energy balance between absorbed UV and reemitted IR radiation is 
approximately satisfied, the IRX--$\Delta \beta$ relation can be derived 
analytically as a function of dust optical depth and dust-star geometry. 
Following \citet{Popping:2017} and \citet{Lin:2021}, we develop a toy 
model parameterized by only these two quantities. 

We assume stellar and dust layers with scale heights $H_*$ and $H_{\rm d}$, 
respectively. The geometry is symmetric, and the dust layer has a uniform 
dust density, $n_{\rm d}$. We define two key parameters: the dust-to-star 
scale-height ratio $R\equiv H_{\rm d}/H_*$, and the fiducial optical depth
\begin{equation}
\tau^{\rm fid}_{\lambda} = \sigma_\lambda n_{\rm d} H_{\rm d},
\end{equation}
where $\sigma_\lambda$ is the dust cross-section, uniquely determined by the 
chosen dust model through its composition and optical properties. We introduce the escape probability at 
wavelength $\lambda$ as
\begin{equation}
P_{{\rm esc}, \lambda} \equiv L_{{\rm obs}, \lambda}/L_{{\rm int}, \lambda},
\end{equation}
where $L_{{\rm obs}, \lambda}$ and $L_{{\rm int}, \lambda}$ are the observed 
and intrinsic luminosities, respectively.

\begin{figure}
    \centering
    \includegraphics[width =\linewidth, clip]{\figdir/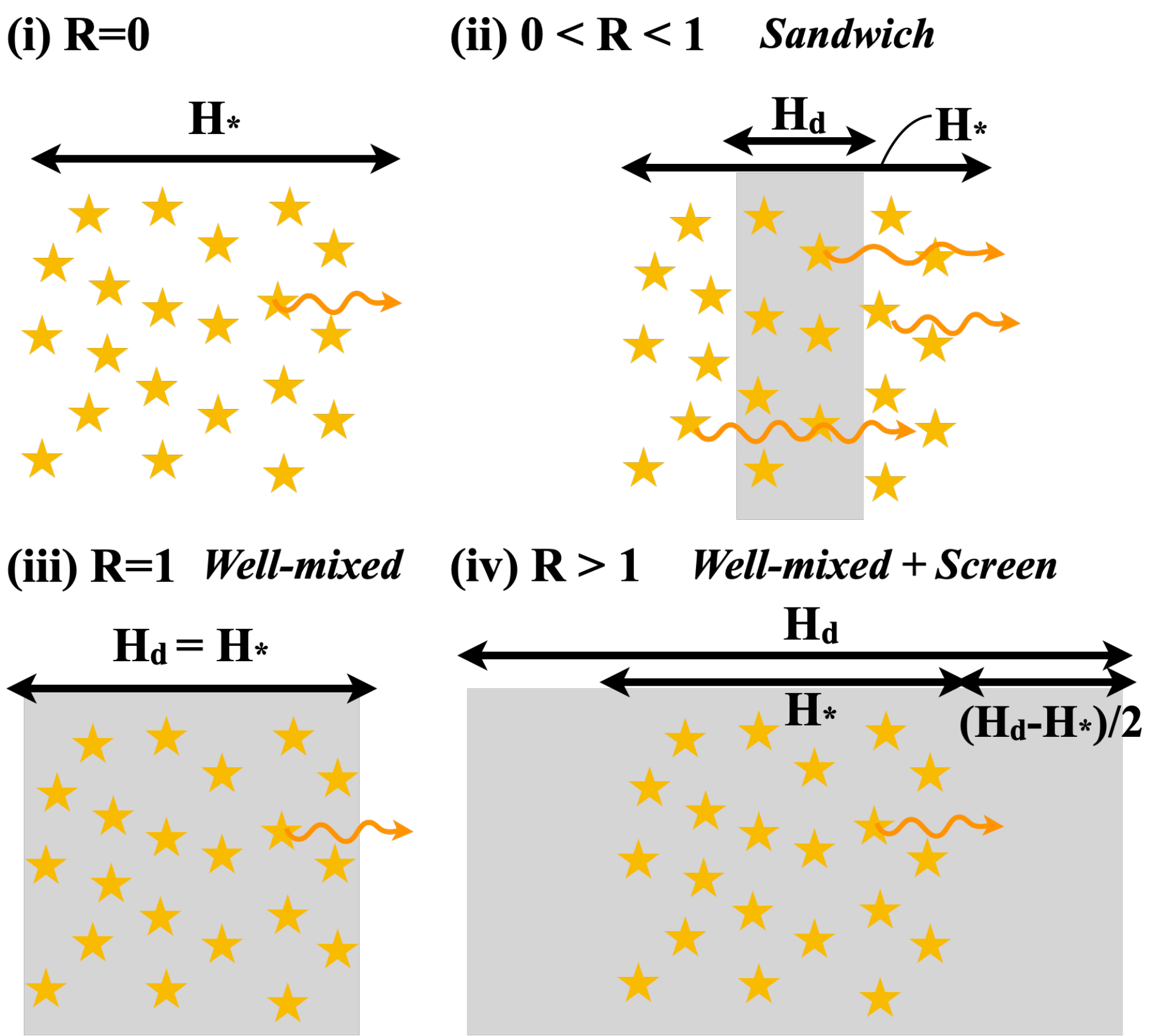}
    \caption{Illustration of the four simplified dust-star geometry models characterized by different dust-to-star scale height ratios ($R \equiv H_{\rm d}/H_*$). Stars represent the stellar layer with scale height $H_*$, and gray shaded areas indicate the uniform dust layer with scale height $H_{\rm d}$. Orange arrows schematically represent escaping UV radiation. The escape probability for each geometry type is derived explicitly in the main text. 
}
    \label{fig:dust_geometry_illustration}
\end{figure}

The dust geometry is categorized into four scenarios based on the scale-height 
ratio $R$, as schematically illustrated in Figure \ref{fig:dust_geometry_illustration}. We assume that these geometries are symmetric about the midplane of the stellar 
distribution.

\noindent{\bf (i) $R = 0$:}
There is no dust layer, and therefore the observed luminosity equals the 
intrinsic luminosity, i.e., $P_{{\rm esc}, \lambda} = 1$.

\noindent{\bf (ii) $0 < R < 1$:}
This corresponds to the {\it sandwich} geometry \citep{Xu_Buat:1995}, with 
an escape probability given by
\begin{equation}
P_{{\rm esc}, \lambda} = \frac{1-R}{2} (1 + e^{-\tau^{\rm fid}_{\lambda} }) 
+ \frac{R}{\tau^{\rm fid}_{\lambda} }(1- e^{-\tau^{\rm fid}_{\lambda} }) .
\end{equation}

\noindent{\bf (iii) $R = 1$:}
This corresponds to a well-mixed geometry, and the escape probability is 
given by
\begin{equation}
P_{{\rm esc}, \lambda} = \frac{1- e^{-\tau^{\rm fid}_{\lambda} }}{\tau^{\rm fid}_{\lambda}}. 
\end{equation}
Cases (ii) and (iii) can be considered as a continuous transition.

\noindent{\bf (iv) $R > 1$:}
This corresponds to a combination of well-mixed and screen geometries. The 
escape probability is then given by
\begin{equation}
P_{{\rm esc}, \lambda} =  \frac{1- e^{-\tau^{\rm mix}_{\lambda} }}{\tau^{\rm mix}_{\lambda}} 
e^{-\tau^{\rm screen}_{\lambda} }, 
\end{equation}
where we define optical depths for the well-mixed and screen components as
\begin{align}
\tau^{\rm mix}_{\lambda} &= \sigma_\lambda n_{\rm d} H_* =  \sigma_\lambda 
n_{\rm d} H_{\rm d} \frac{H_*}{H_{\rm d}} = \frac{\tau^{\rm fid}_{\lambda}}{R}, \\
\tau^{\rm screen}_{\lambda} &= \sigma_\lambda n_{\rm d} \frac{H_{\rm d}- H_*}{2} 
= \frac{\tau^{\rm fid}_{\lambda}}{2} \left(1- \frac{1}{R}\right). 
\end{align}

Next, we derive expressions for IRX and $\beta_{\rm UV}$ from this toy model. 
We introduce an effective optical depth defined as 
$\tau_{{\rm eff}, \lambda} = - \log_{\rm e}(P_{{\rm esc}, \lambda})$. 
Assuming the intrinsic UV luminosity is balanced by the observed UV and 
reemitted IR luminosities, IRX is expressed as
\begin{equation}
{\rm IRX} = \frac{L_{\rm TIR}}{L_{\rm obs, UV}} =\frac{1- e^{-\tau^{\rm eff, UV}}}{e^{-\tau^{\rm eff, UV}}} 
= \frac{1- P_{\rm esc, UV}}{P_{\rm esc, UV}}. \label{eq:IRX_toy_model}
\end{equation}
We adopt a wavelength of 1600 \AA\, as representative of UV radiation.

To define the UV slope, we select two wavelengths consistent with Section 
\ref{subsec:case_study}: $\lambda_1 = 1230$ \AA\, and $\lambda_2 = 3200$ \AA. 
The observed UV slope is then
\begin{align}
\beta_{\rm UV} &= \frac{\log_{10}(f_{\lambda_1})-\log_{10}(f_{\lambda_2})}{\log_{10}(\lambda_1)- \log_{10}(\lambda_2)} \nonumber \\
& = \frac{\log_{10}(f^0_{\lambda_1})-\log_{10}(f^0_{\lambda_2})- (\log_{10}e)\times 
(\tau_{\rm eff, \lambda_1}-\tau_{\rm eff, \lambda_2} )}{\log_{10}(\lambda_1)- \log_{10}(\lambda_2)} \nonumber \\
& = \frac{\log_{10}(f^0_{\lambda_1})-\log_{10}(f^0_{\lambda_2})}{\log_{10}(\lambda_1)- \log_{10}(\lambda_2)} 
+ \frac{-0.43 (\tau_{\rm eff, \lambda_1}-\tau_{\rm eff, \lambda_2} )}{\log_{10}(\lambda_1)- \log_{10}(\lambda_2)} \nonumber \\
& = \beta_{\rm UV, 0} + \Delta \beta, \label{eq:UV_slope_toy_model}
\end{align}
where $f_\lambda$ ($f^0_\lambda$) denotes the observed (intrinsic) flux 
density at wavelength $\lambda$.

Finally, we derive the fiducial optical depth at $\lambda=\lambda_1$ as
\begin{equation}
\tau^{\rm fid}_{\lambda_1} = \tau^{\rm fid}_{\rm UV} \left(\frac{A_{\lambda_1}^{\rm ext}}{A_{\rm UV}^{\rm ext}}\right), 
\end{equation}
where the extinction ratio ${A_{\lambda_1}^{\rm ext}}/{A_{\rm UV}^{\rm ext}}$ is uniquely set by the adopted extinction curve. 
The intrinsic UV slope $\beta_{\rm UV, 0}$ depends on the intrinsic SED, specifically stellar population age (star formation history) and metallicity. According to Eq.~(\ref{eq:UV_slope_toy_model}), $\Delta \beta$ is independent of the intrinsic SED and depends only on the dust geometry, provided that the dust model is fixed. We therefore use $\Delta \beta$ in the following analysis. 

\begin{figure*}
    \centering
    \includegraphics[width =1\linewidth, clip]{\figdir/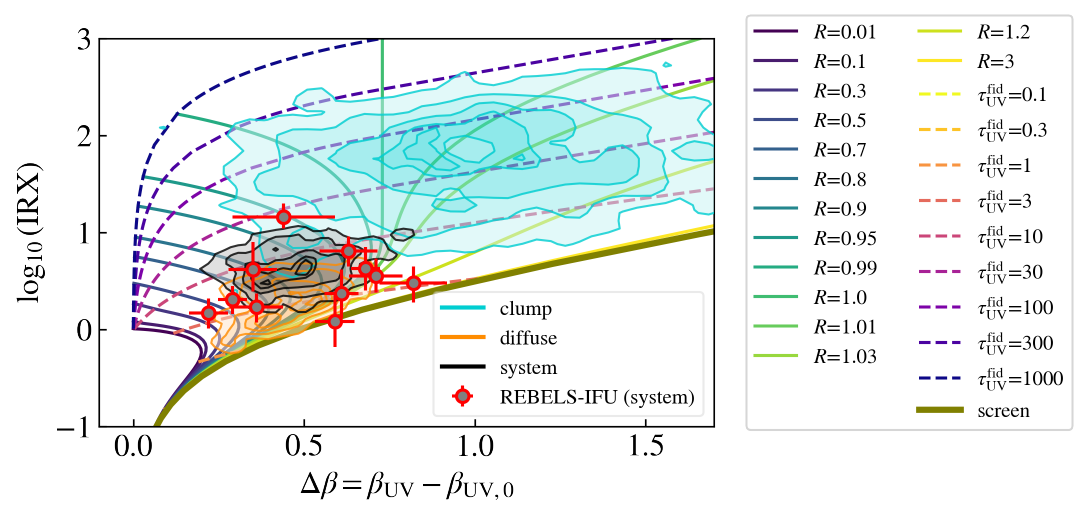}
    \vspace{-5mm}
   \caption{IRX--$\Delta\beta$ relation with toy model predictions, where $\Delta\beta \equiv \beta_{\rm UV} - \beta_{\rm UV,0}$ represents the change in UV slope due to dust attenuation. We show theoretical curves for different combinations of the dust-to-star scale-height ratio ($R$) and fiducial UV optical depth ($\tau^{\rm fid}_{\rm UV}$), which parametrize the dust-star geometry and dust column density, respectively. Each solid curve represents a track of fixed $R$, while each dashed curve represents a track of fixed $\tau^{\rm fid}_{\rm UV}$. The solid 
   olive lines represent the limiting case of a foreground screen geometry ($R \gg 1$). Contours show our simulation results for the clumpy, diffuse, and system-integrated components at $z=6-9$; black for the system-integrated value, cyan for the individual clumps, and orange for the diffuse component. Contours enclose 20\%, 50\%, 80\%, and 95\% of the data. All simulation data assume MW-type dust properties, and the toy model grids are calculated using the MW extinction curve with $R_V = 3.1$ \citep{Cardelli:1989}. For comparison, we plot the REBELS-IFU (system-integrated) data from \citet{Fisher:2025} and \citet{Bowler:2024}. Note that while $\beta_{\rm UV}$ can be measured at clump scales with NIRSpec-IFU observations, IRX and $\Delta \beta$ values are currently only available for system-integrated measurements due to the limited angular resolution of ALMA dust continuum observations and the use of NIRSpec/PRISM data, respectively.
}
    \label{fig:IRX_delta_all_MW}
\end{figure*}

We construct theoretical grids on the IRX--$\Delta \beta$ plane by parameterizing dust-star geometry through $R$ and the fiducial optical depth $\tau^{\rm fid}_{\rm UV}$. Figure~\ref{fig:IRX_delta_all_MW} shows IRX--$\Delta \beta$ grids obtained by varying these two parameters.

In the case of $R=0$, the position on the IRX--$\Delta \beta$ plane is fixed at $(\Delta \beta,\, {\rm IRX})=(0,\, 0)$. As $R \rightarrow 0$, the toy model represents a scenario with a very thin dust sheet placed symmetrically at the center of the stellar distribution. In the optically thick regime ($\tau^{\rm fid}_{\rm UV} \gg 1$), radiation from behind this central dust sheet is absorbed and reemitted in the IR, whereas radiation from the front side escapes directly as UV emission. These two contributions become equal, causing 
IRX to asymptotically approach unity. This IRX $\sim$ 1 limit represents a limitation of the idealized symmetric geometry and is not realized in actual clumps. For $0 < R < 1$, the grid curves exhibit a characteristic inverted-C shape because escaping radiation includes reddened contributions from the obscured backside. When $R = 1$ (fully well-mixed), the UV slope remains nearly constant at $\Delta \beta \simeq -\log_{\rm e}(\tau^{\rm fid}_{\lambda_1}/\tau^{\rm fid}_{\lambda_2})/\log_{\rm e}(\lambda_1/\lambda_2)$ 
as the optical depth increases ($\tau^{\rm fid}_{\rm UV}\gtrsim1$), 
and IRX increases vertically. For large $R$ values ($R \gtrsim 3$), the geometry approaches a screen model as shown by the solid black line.

In simulations, intrinsic physical properties such as the intrinsic UV slope 
($\beta_{\rm UV,0}$) are readily available for each component, and the dust mixture can be consistently fixed (MW-type dust is assumed throughout 
Figure \ref{fig:IRX_delta_all_MW}). In what follows, we therefore utilize the 
IRX--$\Delta \beta$ plane to interpret the dust geometry in our simulated galaxies.

Figure~\ref{fig:IRX_delta_all_MW} shows our simulated galaxies on the IRX--$\Delta\beta$ plane, enabling direct interpretation of dust-star geometry\footnote{We note that our toy model assumes component-wise energy balance, which strictly holds only when integrated over the entire solid angle and not necessarily for a single line of sight. This assumption is well-justified for compact clumps with small surface areas, where external UV contributions are negligible. For extended diffuse components, deviations from strict energy balance may occur, though the qualitative trends remain valid.}. From this figure, we find that clumps occupy the region with 
%$\tau_{\rm UV}^{\rm fid} \gtrsim 10$ 
$\tau_{\rm UV}^{\rm fid} \sim 10-300$ and $R \simeq 1$, indicating well-mixed geometry. System-integrated values have $\tau_{\rm UV}^{\rm fid} \sim 3$--$20$ and $R \simeq 0.8$, exhibiting sandwich geometries. Diffuse components have small fiducial optical depths ($\tau_{\rm UV}^{\rm fid} \lesssim 10$), making geometry determinations difficult since $\Delta \beta$ takes similar values regardless of $R$.

In terms of optical depth (i.e., dust column density), clumps have $\sim$10 times larger column densities than the system-integrated values (with extreme cases up to $\sim$100 times). Despite their large optical depths, the escape probability for well-mixed geometry scales as $P_{{\rm esc}, {\rm UV}} \propto {\tau^{\rm fid}_{\rm UV}}^{-1}$, allowing a small fraction of UV photons to escape. These high optical depths result from gas compression during interactions with companion galaxies, significantly enhancing local dust column densities.

We also plot observational results of system-integrated $z\sim 7$ galaxies obtained from the REBELS-IFU survey as gray circles in Figure~\ref{fig:IRX_delta_all_MW}. We adopt IRX results from \citet{Bowler:2024} and $\Delta \beta$ from \citet{Fisher:2025}. Here, $\Delta \beta$ is obtained using flexible attenuation curve fitting, which simultaneously constrains the intrinsic stellar SED and the attenuation curve shape. These data points are located in similar regions to the system-integrated values in our simulation results.

\begin{figure*}
    \centering
    \includegraphics[width =\linewidth, clip]{\figdir/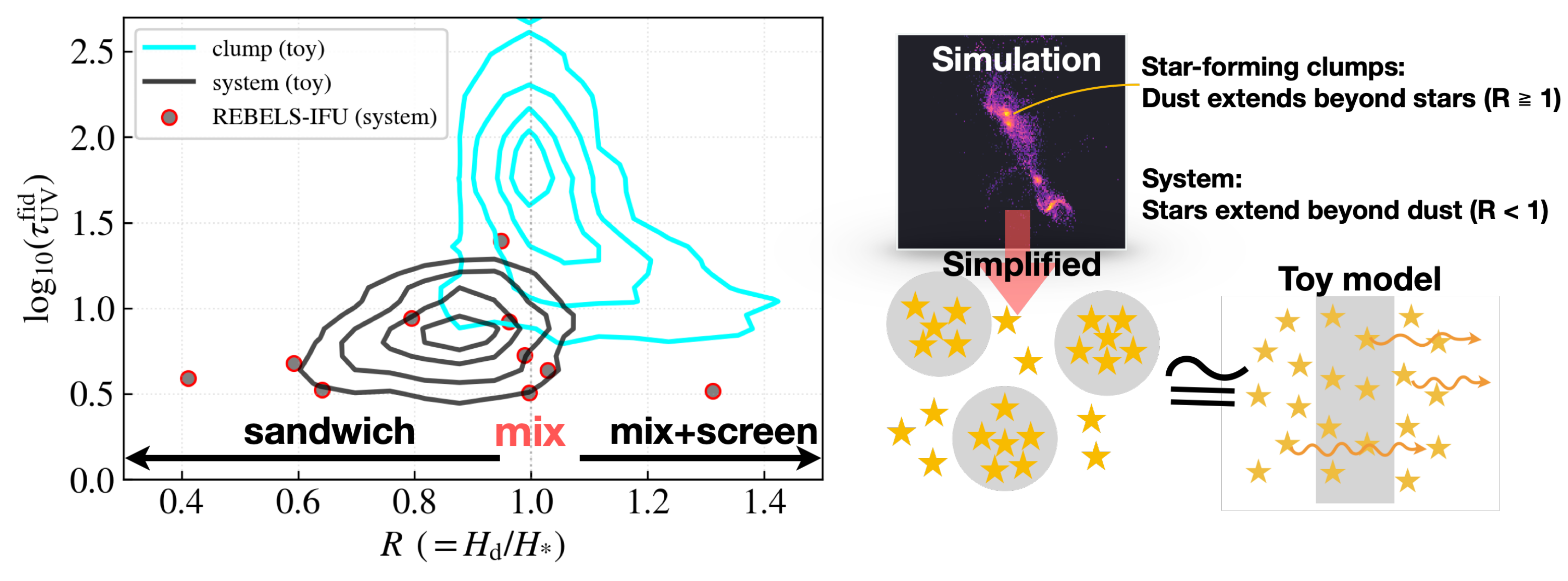}
    \vspace{-6mm}
    \caption{Left: Distributions of the dust-to-star scale-height ratio ($R$) and fiducial UV optical depth ($\tau^{\rm fid}_{\rm UV}$) for clump components (cyan) and system-integrated values (black). 
These contours and the gray circles show values inferred from the toy-model grid matching in Figure \ref{fig:IRX_delta_all_MW}, where the latter represents the REBELS-IFU (system-integrated) data. The contours enclose 20\%, 50\%, 80\%, and 95\% of the data.   
    Right: Schematic illustration connecting the spatial structure of simulated galaxies to the toy model framework: clumpy regions exhibit dust-extended or co-spatial dust/stellar distributions 
    ($R \geq 1$)
    with $\sim$1 dex larger dust column densities than system-averaged, while system-integrated values represent the stellar-extended geometry due to diffuse components ($R<1$).
    }
    \label{fig:schematic}
\end{figure*}

%To validate the toy-model predictions, we also directly calculate $R$ and $\tau_{\rm UV}^{\rm fid}$ from the simulations. This direct calculation is feasible only for clump components, as defining $R$ for diffuse regions and system-integrated values is ambiguous. For clumps, we define the scale heights for dust and stars as the half-mass radii within each clump. The left panel of Figure~\ref{fig:schematic} shows the distributions of $R$ and $\tau_{\rm UV}^{\rm fid}$ for each component. The directly measured $R$ values for clumps exhibit larger scatter but have a median of $R \simeq 1$, with the scatter systematically extending toward $R > 1$, consistent with the toy-model analysis. Clumps have high dust column densities with the median value $\log_{10}\tau^{\rm fid}_{\rm UV} \sim 1.6\,(1.9)$ for toy-model-inferred (directly measured) values. The right panel of Figure~\ref{fig:schematic} shows a schematic illustration of our results: star-forming clumps ($\sim$100~pc scale) have $\sim$1~dex larger column densities than the system average and exhibit co-spatial or dust-extended geometries. System-integrated values, however, represent a mixture of clumps and weakly attenuated diffuse components, which is interpreted as a sandwich geometry (i.e., star-extended geometry) in the 1D toy model. We also estimate $R$ and $\tau_{\rm UV}^{\rm fid}$ for the REBELS-IFU samples in the same manner, finding that they are consistent with the system-integrated values in our simulations ($R \lesssim 1$, $\log_{10}\tau_{\rm UV}^{\rm fid}\simeq 0.5$--$1.5$).

To interpret the dust-star geometry more quantitatively, we extract the 
corresponding ($R$, $\tau_{\rm UV}^{\rm fid}$) combinations from their 
positions on the IRX--$\Delta\beta$ plane. The left panel of 
Figure~\ref{fig:schematic} shows how different components map onto the 
toy-model parameter space. Clumps have a median of $R = 1.0$, with scatter 
systematically extending toward $R > 1$, and high optical depths with median 
$\log_{10}\tau^{\rm fid}_{\rm UV} \sim 1.6$ (5th-95th percentile:1.0-2.2). System-integrated values have 
$R < 1$ with median $\log_{10}\tau^{\rm fid}_{\rm UV} \sim 0.85$ (5th-95th percentile:0.59-1.15).

To validate these toy-model-inferred values, we directly calculated the 
dust-star geometry parameters from our simulations. While defining scale 
heights for extended components is challenging, we measured the half-mass 
radius ratio between dust and stars for compact clumps, finding values 
of $r_d/r_* \sim 1$ (corresponding to $R \sim 1$), consistent with the 
well-mixed interpretation. The directly calculated fiducial optical depths for clumps, derived from dust column density maps, span $\log_{10} \tau^{\rm fid}_{\rm UV} \sim 1.5$--$2.4$ (median $\sim 1.9$), in reasonable agreement with the toy-model-inferred median of $\sim 1.6$.
This consistency validates that the toy model successfully captures the dust-star geometry of our simulated galaxies.

We also apply this framework to the observed REBELS-IFU galaxies 
(system-integrated) and find that most samples span $R \lesssim 1$ and 
$\log_{10} \tau^{\rm fid}_{\rm UV} \sim 0.5$--$1.5$, consistent with the 
system-integrated values in our simulated galaxies. The right panel of 
Figure~\ref{fig:schematic} schematically illustrates these results: 
star-forming clumps ($\sim$100~pc scale) have $\sim$1~dex larger column 
densities than the system average and exhibit well-mixed geometries, while 
system-integrated values represent a mixture of clumps and weakly attenuated 
diffuse components, interpreted as sandwich geometry in the 1D toy model.

In summary, given a known dust model (extinction curve), the IRX--$\Delta \beta$ plane effectively diagnoses dust-star geometry, distinguishing between the effects of optical depth and geometric configuration. Recent JWST observations have enabled the derivation of attenuation curves, which will allow future observations to derive intrinsic UV slopes and $\Delta \beta$, ultimately enabling quantitative decomposition of these effects using this diagnostic plane.

\subsection{Implications for Spatially Resolved Observations} \label{subsec:implications}
Our spatially resolved analysis of dust absorption and reemission properties reveals systematic variations that have important implications for interpreting observations of high-redshift galaxies.

First, we find that dust temperatures vary spatially (Figure \ref{fig:dust_properties_statistical}), with clump peak temperatures up to $\Delta T_{\rm d, peak} \sim 20$ K higher than system-integrated values. Such temperature variations reflect differences in local heating sources and dust column densities. While direct spatially resolved dust temperature measurements require high-resolution multi-band IR observations that remain challenging \citep{Villanueva:2024}, alternative methods using spatially resolved UV continuum with single-band IR observations \citep{Inoue:2020, Fudamoto:2023} or \CII emission lines \citep{Sommovigo:2021} offer promising avenues for future studies. Such measurements would enable direct comparison with our predictions and provide constraints on the spatial distribution of heating sources (e.g., young stellar populations, evolved stars, AGN) and dust masses within high-redshift galaxies.

Second, variations in attenuation curves directly affect pixel-by-pixel SED fitting (Figure \ref{fig:attenuation_curve}). Recent JWST NIRCam observations have enabled pixel-by-pixel SED fitting of high-redshift galaxies \citep[e.g.,][]{Gimenez-Arteaga:2023, Gimenez-Arteaga:2024}. However, these studies typically assume a fixed attenuation law (often 
Calzetti) for all pixels\footnote{Just recently, \citet{Markov:2026} conducted pixel-by-pixel SED fittings using a flexible attenuation law.}. Our results show that star-forming clumps have grayer attenuation curves than the Calzetti law, while diffuse regions (tails and bridges) have steeper curves.

If a fixed Calzetti law is applied to all pixels, systematic biases arise. For clumps with grayer curves, the Calzetti law (which is steeper) will underestimate $A_{\rm V}$. To reproduce the observed UV colors with lower $A_{\rm V}$, SED fitting will favor younger stellar populations with intrinsically bluer colors and lower mass-to-light ratios, leading to 
underestimated stellar masses. Conversely, for diffuse components with steeper curves, the Calzetti law (which is shallower) will overestimate $A_{\rm V}$, favoring older populations with higher mass-to-light ratios and leading to overestimated stellar masses.

While the impact of attenuation curve variations on integrated galaxy stellar mass estimates has been studied \citep[e.g.,][]{Lo_Faro:2017, Pforr:2012}, our work is the first to investigate these effects on spatially resolved scales within individual galaxies. Recent studies have begun using flexible attenuation curve models to characterize system-integrated properties in JWST observations \citep{Markov:2025, Shivaei:2025, Fisher:2025} and deriving intrinsic UV slopes \citep{Fisher:2025}. Our results suggest that such flexible parametrizations should also be applied at the pixel-by-pixel level, allowing attenuation curves to vary spatially within individual galaxies rather than adopting a single curve for the entire system.

Such spatially varying attenuation curves would enable the derivation of intrinsic UV slopes at each spatial location. Combined with spatially resolved IRX measurements, this would allow us to construct spatially resolved IRX--$\Delta \beta$ maps and thereby probe the dust-star geometry and optical depth on sub-galactic scales. Although only a handful of studies have derived spatially resolved IRX for $z \gtrsim 5$ galaxies \citep{Sugahara:2025, Lines:2025, Mawatari:2026, Bakx:2025}, we expect such observations to become increasingly available with future JWST/ALMA programs.

\subsection{Caveats}\label{subsec:caveats}
There are several caveats to our study. First, we adopt a fixed dust–to–metal ratio of 0.4. According to \citet{Dayal:2022}, DTM keeps the constant value of 0.37 when considering only stellar-path production. However, in reality, processes such as astration and destruction act to reduce the DTM, while ejection (which preferentially removes gas-phase metals) and grain growth act to increase it \citep[e.g.,][]{Popping:2017, Aoyama:2017, Esmerian:2024, Choban:2025, Toyouchi:2025}. Moreover, the efficiency of grain growth depends on the accretion timescale for gas-phase metals to stick onto dust grains, which scales as $t_{\rm acc}\propto n^{-1}T^{-1/2}Z^{-1}$. This equation implies that higher-density, lower-temperature, higher-metallicity environments such as molecular clouds favor collisions and subsequent sticking between gas-phase metals and grains \citep{Dwek:1998, Zhukovska:2008}. 
In future work we will therefore implement a fully time-dependent DTM obtained from semi-analytic models \citep[e.g.,][]{Dayal:2022, Tsuna:2023, Toyouchi:2025, Toyouchi:2026} and cosmological simulations \citep[e.g.,][]{Aoyama:2017, Graziani:2020, Kannan:2025}. 

Second, we assume a fixed dust composition and size distribution: MW– and SMC-like mixtures. High-redshift dust is believed to form predominantly in core-collapse supernovae \citep{Mancini:2015, DellAgli:2019, Lesniewska:2019, Schneider_Maiolino:2024} and to be processed by shocks in such a way that small grains are preferentially destroyed, biasing the population toward larger particles \citep{Makiya:2022, Narayanan:2026}. Because grains with radii exceeding the photon wavelength scatter and absorb 
nearly independently of wavelength \citep{Mie:1908}, a 
distribution dominated by large grains produces a gray (flat) extinction curve. Exploring evolving dust yields, shattering, coagulation, and ISM growth in a self-consistent framework that allows both composition and size distribution to vary with redshift is a natural next step for future investigations.

Third, we identify clumps based on SFR surface densities, a method that physically targets gravitationally bound structures. In practice, however, the definition of a ``clump’’ and the choice of aperture size vary significantly across observational studies. If an observational aperture is larger than our model-defined clump scale, it may incorporate the surrounding diffuse component, resulting in a lower IRX that approaches galaxy-integrated values. Therefore, our results should be viewed as a conservative characterization of the intrinsic properties of the clump structures themselves, representing the high-density limit of these systems. Creating realistic mock observations that include noise and PSF effects, and identifying clumps based on flux rather than SFR density \citep[e.g.,][]{Punyasheel:2025, Ceverino:2026}, would facilitate a more direct comparison with observations.

\section{Summary and Conclusions}\label{sec:conclusion}
We present spatially resolved dust observables, including dust temperature, UV slope, attenuation curves, and infrared excess (IRX), for clumpy galaxies ($M_* \simeq 10^{8-10}\, M_\odot$) at $z= 6$--$9$ using cosmological zoom-in simulations. By performing postprocessing dust radiative transfer calculations, we obtain pixel-by-pixel SEDs for each simulated galaxy. We identify star-forming giant clumps ($R \gtrsim 100\, {\rm pc}$, ${\rm SFR} \gtrsim 1\, M_\odot\, {\rm yr^{-1}}$) based on SFR surface density, detecting 376 clumpy systems and 1059 individual clumps in total at $z= 6$--$9$. For each clumpy system, we analyze three components: clumps, diffuse regions, and system-integrated values, and statistically compare their dust observables.

We find that clumps exhibit systematically higher $T_{\rm d, peak}$, $A_{\rm V}$, $A_{\rm UV}$, and redder $\beta_{\rm UV}$ compared to system-integrated values, with large scatter. In contrast, diffuse components show similar $T_{\rm d, peak}$ and $\beta_{\rm UV}$ but smaller $A_{\rm V}$ and $A_{\rm UV}$ than system-integrated values, with much smaller scatter. Some clumps have temperatures up to 20~K higher than their host galaxies' system-integrated values. The median IRX of clumps is 1--2~dex larger than both diffuse and system-integrated values, indicating dramatically different dust obscuration levels within individual galaxies.

For attenuation curves, we find that system-integrated values for our simulated galaxies are grayer than the Calzetti law. For spatially resolved results, we find that clumps have grayer attenuation curves than the system-integrated values, while diffuse components exhibit much steeper curves due to enhanced scattering effects in optically thin regions. Even though our postprocessing dust model assumes MW-like (or SMC-like) dust, we obtain grayer curves, implying that optical depth and dust-star geometry are the main contributors to this shallowness.

To decompose these two factors, we develop a toy model that utilizes the IRX--$\Delta\beta$ plane, where $\Delta \beta$ is the difference between the attenuated and intrinsic UV slopes. This model characterizes dust observables using two parameters: dust optical depth ($\tau^{\rm fid}_{\rm UV}$, i.e., dust column density) and the dust-to-star scale-height ratio ($R$). We find that clumps have $\sim$10 times higher dust column densities than system-integrated values and exhibit co-spatial or dust-extended geometries, while system-integrated values show stellar-extended geometries due to contributions from optically thin diffuse components. Comparison with REBELS-IFU observations shows 
that observed $z \sim 7$ galaxies have dust properties consistent with those of our system-integrated values.
%We also directly calculate the scale-height ratio and optical depth for clumps from the simulations and validate our toy model predictions.

These results have important implications for interpreting spatially resolved observations of high-redshift galaxies. Pixel-by-pixel measurements of attenuation curves will be crucial for accurately determining physical properties from SED fitting. Furthermore, as clump-by-clump IRX--$\Delta \beta$ measurements become available, they will enable direct constraints on dust-star geometry and optical depth. Such spatially resolved observations are becoming feasible, particularly in gravitationally lensed systems (e.g., CANUCS (JWST GTO program), VENUS survey: JWST GO-6882). While our analysis focuses on high-redshift clumpy galaxies, the methodology and findings are equally applicable to spatially resolved observations at lower redshifts (e.g., HIDING survey, ALMA Cycle 12 large program).

The combination of next-generation spatially resolved observations from JWST and ALMA with detailed radiative transfer modeling will be essential for understanding the physical conditions in individual star-forming regions and for accurately characterizing dust properties across cosmic time.

\section{Acknowledgements}
We are grateful to the anonymous referee for providing valuable comments that have greatly improved the paper.
 YN thanks Masato Hagimoto, Amiel Sternberg, and Rachel Somerville, Irene Shivaei, Masami Ouchi, Kartheik Iyer, and Laura Sommovigo for frutiful discussion. This work made use of v2.3 of the Binary Population and Spectral Synthesis (BPASS) models as described in \citet{Stanway:2018} and \citet{Byrne:2022}. Numerical computations and analyses were carried out on the Cray XD2000 and the analysis servers at Center for Computational Astrophysics, National Astronomical Observatory of Japan. YN acknowledges funding from JSPS KAKENHI Grant Number 23KJ0728, a JSR fellowship, and Flatiron Research Fellowhship. The Flatiron Institute is a division of the Simons Foundation. AKI is supported by JSPS KAKENHI Grant Number 23H00131. TH acknowledges financial support from the JSPS (19KK0353) and the Kyoto University Foundation. DC is supported by research grant PID2021-122603NB-C21 funded by the Ministerio de Ciencia, Innovación y Universidades (MI-CIU/FEDER), project PID2024-156100NB-C21 financed by MI-CIU/AEI /10.13039/501100011033 / FEDER, EU., and the research grant CNS2024-154550 funded by MI-CIU/AEI/10.13039/501100011033.

\appendix 
\counterwithin{figure}{section}
\counterwithin{table}{section}
\section{numerical convergence and additional physical processes} \label{sec:numerical_convergence}
In this appendix, we check the numerical convergence of our RT simulations and the effects of stochastic heating and self-absorption. Table \ref{table:convergence_test} summarizes our test calculation. Here, $n_{\rm p}$ denotes the number of photon packets launched {by the source}, and $n_{\lambda}$ denotes the number of grid points used for the wavelength grid. By default, we use $(n_{\rm p}, n_{\rm \lambda})=(10^7, 150)$.

\begin{figure}
    \centering
    \includegraphics[width =\linewidth, clip]{\figdir/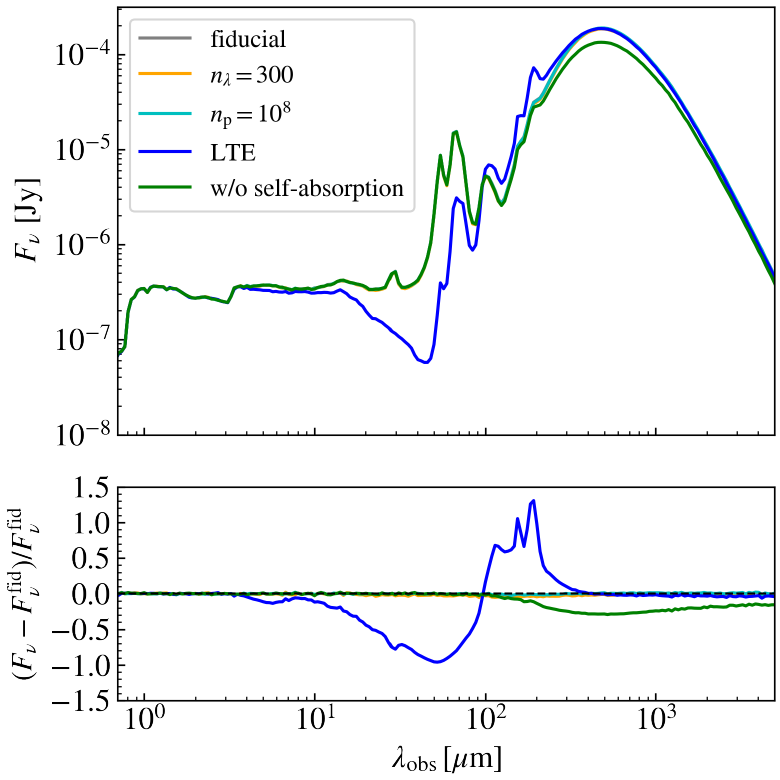}
    \caption{SEDs of FL957 at $z=7.7$ computed with different SKIRT settings. In the upper panel, we plot the SED for the fiducial case (gray) and two runs with more wavelength bins (orange, $n_\lambda = 300$), and more photon packages (light blue, $n_{\rm p} = 10^8$) launched per wavelength bin. Additionally, we show the outcome of simulations for which stochastic heating or self-absorption is turned off (blue: ``LTE’’, green:``w/o self-absorption’’). The detailed SKIRT settings in each model are described in Table \ref{table:convergence_test}. In the lower panel, we show the fractional difference $(F_\nu - F_\nu^{\rm fid})/F_\nu^{\rm fid}$ between the various runs and the fiducial model.
}
    \label{fig:convergence_test}
\end{figure}

In Figure \ref{fig:convergence_test}, we show the SEDs for 
$(n_{\rm p}, n_{\rm \lambda})=(10^8, 150)$ and $(10^7, 300)$ for FL957 at $z=7.7$, the same snapshot shown in Figure \ref{fig:projection_FL957_a0p115}. The differences from the default case are less than 2 percent across most wavelength regions in the rest-UV and rest-FIR. Such small differences are negligible when integrated over observed 
photometric filters, confirming that our results are numerically converged. 

We also check the convergence of the derived physical quantities ($\beta_{\rm UV}, \, \log_{10}({\rm IRX}), \, A_{\rm UV},\, {A_{\rm V}}$, and $S\equiv A_{\rm UV}/A_{\rm V}$) for each component of FL957 at $z=7.7$, as summarized in Table \ref{table:convergence_test_component}. For the system-integrated diffuse components, all quantities agree to within 4\%. For clump components, $\beta_{\rm UV}$ can vary by up to 0.3, while $\log_{10}({\rm IRX}), \,A_{\rm UV}, \, A_{\rm V},$ and $S$ agree to within 0.03 dex, 5\%, 7\%, and 20\%, respectively. The largest deviations are found in clump 2, which has a low signal-to-noise ratio due to its small area. These results confirm that our fiducial settings do not significantly affect out conclusions.

Additionally, we also plot a case without stochastic heating, i.e., the assumption that all grains are in thermal equilibrium with the incident radiation field. The SED plotted in blue lacks the rest-frame mid-IR emission produced from very small grains such as PAHs. They underestimate reemission flux from stochastic heating by up to a factor of two. This rest-frame mid-IR windows overlaps well with the spectral range of the proposed PRIMA instrument \citep{Moullet:2023, Moullet:2025}, making it a crucial probe for distinguishing the effects of stochastic heating in these high-$z$ galaxies.

The green line is an SED without self-absorption of dust, and it shows a net decrease of flux both in the mid-IR and FIR. Without self-absorption, the reemission from dust continuum absorption within optically thick regions is not accounted for, leading to an underestimation of IR flux. Consequently, the flux can be underestimated by up to 25\% with respect to the fiducial simulation. 

\begin{table}[]
\centering
\begin{tabular}{l|cccc} \hline
name    & $n_{\rm p}$     & $n_{\lambda}$ & \shortstack{self-\\absorption}  & \shortstack{stochastic\\heating} \\ \hline\hline
fiducial & $10^7$         & 150         & $\bigcirc$        & $\bigcirc$          \\
nw300   & $10^7$          & 300         & $\bigcirc$        & $\bigcirc$           \\
np1e8   & $10^8$          & 150         & $\bigcirc$        & $\bigcirc$           \\
LTE     & $10^7$          & 150         & $\bigcirc$        & $\times$              \\
w/o self-absorption & $10^7$   & 150    & $\times$          & $\bigcirc$           \\ \hline       
\end{tabular} 
\caption{The parameters used in each SKIRT calculation in Appendix \ref{sec:numerical_convergence}. All cases adopt MW-like dust model \citep{Weingartner_Draine:2001} and fixed dust-to-metal ratio of 0.4. Here, $n_{\rm p}$ is the number of photon packets launched per wavelength grid point, and $n_{\lambda}$ is the number of wavelength grid points.}\label{table:convergence_test}
\end{table}

\begin{table}[ht]
\centering
\renewcommand{\arraystretch}{1.2}
\begin{tabular}{@{}llcccccccc@{}}
\hline
Component & Case & $\beta_{\rm UV}$ & $\log({\rm IRX})$ & $A_{\rm UV}$ & $A_{\rm V}$ & $S$ \\
\hline\hline
\multirow{3}{*}{clump 1} & Default & $-1.68$ & $1.99$ & $4.6$ & $2.84$ & $1.62$ \\
 & $n_\lambda=300$ & $-2.03$ & $2.01$ & $4.63$ & $2.97$ & $1.57$ \\
 & $n_{\rm p}=10^8$ & $-1.73$ & $2.00$ & $4.63$ & $2.90$ & $1.60$ \\
\hline
\multirow{3}{*}{clump 2} & Default & $-2.30$ & $0.99$ & $1.82$ & $0.97$ & $1.89$ \\
 & $n_\lambda=300$ & $-2.56$ & $0.71$ & $1.82$ & $0.97$ & $1.89$ \\
 & $n_{\rm p}=10^8$ & $-2.45$ & $0.70$ & $1.77$ & $1.03$ & $1.72$ \\
\hline
\multirow{3}{*}{clump 3} & Default & $-2.49$ & $0.94$ & $2.21$ & $1.58$ & $1.40$ \\
 & $n_\lambda=300$ & $-2.34$ & $0.93$ & $2.24$ & $1.58$ & $1.42$ \\
 & $n_{\rm p}=10^8$ & $-2.36$ & $0.95$ & $2.25$ & $1.54$ & $1.46$ \\
\hline
\multirow{3}{*}{clump 4} & Default & $-2.11$ & $1.29$ & $2.95$ & $1.68$ & $1.76$ \\
 & $n_\lambda=300$ & $-2.05$ & $1.31$ & $2.98$ & $1.73$ & $1.71$ \\
 & $n_{\rm p}=10^8$ & $-2.13$ & $1.31$ & $2.98$ & $1.69$ & $1.76$ \\
\hline
\multirow{3}{*}{diffuse} & Default & $-2.35$ & $0.07$ & $0.64$ & $0.10$ & $6.79$ \\
 & $n_\lambda=300$ & $-2.35$ & $0.06$ & $0.64$ & $0.10$ & $6.33$ \\
 & $n_{\rm p}=10^8$ & $-2.35$ & $0.07$ & $0.64$ & $0.10$ & $6.59$ \\
\hline
\multirow{3}{*}{system} & Default & $-2.34$ & $0.34$ & $1.02$ & $0.38$ & $2.59$ \\
 & $n_\lambda=300$ & $-2.34$ & $0.34$ & $1.02$ & $0.39$ & $2.61$ \\
 & $n_{\rm p}=10^8$ & $-2.34$ & $0.34$ & $1.02$ & $0.39$ & $2.69$ \\
\hline
\end{tabular}
\caption{Convergence test for the derived physical quantities for each component of FL957 at $z=7.70$. The default case uses $(n_{\rm p}, n_{\rm \lambda}) = (10^7, 150)$.}
\label{table:convergence_test_component}
\end{table}

\begin{comment}
\section{The effect of inclination} \label{sec:inclination_test}
Since different lines-of-sight (LoS) can result in different observed SEDs, we reran our fiducial simulation using 8 additional LoS. Figure \ref{fig:inclination_test} shows the SEDs for a merger-induced clumpy galaxy FL957 at $z=7.7$ (the same galaxy as in Appendix \ref{sec:numerical_convergence}). The effect of inclination on the 
SEDs is relatively minor, with a maximum deviation of 36\% in the 
rest-UV and a median deviation of 12\%. This modest variation reflects the three-dimensional distribution of clumps in merger-driven galaxies, which are not confined to a single plane.

\begin{figure}
    \centering
    \includegraphics[width =\linewidth, clip]{\figdir/inclination_test.pdf}
    \caption{The top panel shows the SED for the clumpy galaxy FL957 at $z=7.7$, comparing our fiducial model (xy-plane, black) with 8 other lines-of-sight (LoS, green lines). The black dashed line shows the intrinsic SED. The bottom panel shows the fractional difference $(F_\nu - F_{\rm fid})/F_{\rm fid}$ between each LoS and the fiducial xy-plane.}
    \label{fig:inclination_test}
\end{figure}
\end{comment}
\section{Physical Properties of the Clumpy Galaxy FL957 at $z=7.7$} 
\label{ref:table_FL957}
In Section~\ref{subsec:case_study}, we present a detailed case study of FL957 at $z=7.7$. Tables~\ref{table:FL957_physical_properties} and \ref{table:FL957_dust_properties} summarize the physical and dust properties for each component (clumps, diffuse, and system-integrated) of this galaxy.
\begin{table*}[ht]
\centering
\renewcommand{\arraystretch}{1.2}
\begin{tabular}{lcccccccc}
\hline
\multirow{2}{*}{Component} 
 & $M_{\rm *,all}$ 
 & $M_{\rm *,young}$ 
 & Age$_{\rm mw}$ 
 & $\langle n_{\rm gas}\rangle$ 
 & $\Sigma_{\rm SFR}$ 
 & SFR 
 & sSFR 
 & $R_{\rm clump}$ \\
 & [M$_\odot$] 
 & [M$_\odot$] 
 & [Myr] 
 & [cm$^{-3}$] 
 & [M$_\odot$ yr$^{-1}$ kpc$^{-2}$] 
 & [M$_\odot$ yr$^{-1}$] 
 & [Gyr$^{-1}$] 
 & [pc] \\
\hline \hline
Clump 1 & $3.21\times10^{8}$ & $8.05\times10^{7}$ & 39 & 492 & 115 & 8.1 & 25.1 & 149 \\
Clump 2 & $3.63\times10^{7}$ & $2.37\times10^{7}$ & 26 & 345 & 59 & 2.4 & 65.2 & 113 \\
Clump 3 & $1.93\times10^{8}$ & $4.63\times10^{7}$ & 34 & 191 & 71 & 4.6 & 24.0 & 144 \\
Clump 4 & $1.32\times10^{8}$ & $8.26\times10^{7}$ & 30 & 400 & 107 & 8.3 & 62.7 & 157 \\
Diffuse & $2.72\times10^{9}$ & $4.25\times10^{8}$ & 100 & 1.9 & - & 42.5 & 15.6 & -- \\
System  & $3.40\times10^{9}$ & $6.58\times10^{8}$ & 87 & 2.8 & - & 65.8 & 19.4 & -- \\
\hline
\end{tabular}
\caption{Summary of physical properties of FL957 at $z=7.70$, which contains four clumps within a $10 \times 10$ kpc$^2$ region. $M_{*, {\rm young}}$ refers to stellar masses younger than 10 Myr, and $M_{*, {\rm all}}$ refers to stellar masses for all ages. Clump age (${\rm Age_{mw}}$) is calculated as the mass-weighted stellar age within the clump. The clump radius $R_{\rm clump}$ is obtained as $R_{\rm clump} = \sqrt{\Delta^2 N_{\rm grid}/\pi}$, where $\Delta$ is the pixel size and $N_{\rm grid}$ is the number of pixels within the clump. The values of $\Sigma_{\rm SFR}$ and $R_{\rm clump}$ for the diffuse and system-integrated cases are not reported because both cases include pixels with no young stars ($<$10 Myr), making a meaningful definition of the area difficult. 
} \label{table:FL957_physical_properties}
\end{table*}

\begin{table*}[h]
\centering
\renewcommand{\arraystretch}{1.2}
\begin{tabular}{lccccccccc}
\hline 
Component & $\beta_{\rm UV}$ & $\beta_{\rm UV,0}$ & $\log_{10}({\rm IRX})$ 
& $L_{\rm UV,1600}$ [L$_\odot$] & $L_{\rm IR}$ [L$_\odot$] 
& $T_{\rm d, peak}$ [K] & $A_{\rm UV}$ & $A_V$ & $S=A_{\rm UV}/A_V$ \\
\hline \hline
Clump 1 & -1.68 & -2.76 & 1.99 & $6.44\times10^{8}$  & $6.33\times10^{10}$ & 55.0 & 4.60 & 2.84 & 1.62 \\
Clump 2 & -2.30 & -2.80 & 0.68 & $2.23\times10^{9}$  & $1.07\times10^{10}$ & 57.4 & 1.75 & 1.12 & 1.56 \\
Clump 3 & -2.49 & -2.78 & 0.94 & $3.48\times10^{9}$  & $3.01\times10^{10}$ & 52.7 & 2.21 & 1.58 & 1.40 \\
Clump 4 & -2.11 & -2.85 & 1.29 & $2.69\times10^{9}$  & $5.26\times10^{10}$ & 52.7 & 2.95 & 1.68 & 1.76 \\
Diffuse  & -2.35 & -2.70 & 0.07 & $1.33\times10^{11}$ & $1.56\times10^{11}$ & 48.3 & 0.64 & 0.09 & 6.79 \\
System   & -2.34 & -2.74 & 0.34 & $1.41\times10^{11}$ & $3.11\times10^{11}$ & 52.7 & 1.02 & 0.38 & 2.69 \\
\hline
\end{tabular}
\caption{Summary of dust observables for FL957 at $z=7.70$ (same system as in 
Table \ref{table:FL957_physical_properties}). Dust observables are defined 
as in Table \ref{table:statistical_dust_properties}.
}\label{table:FL957_dust_properties}
\end{table*}

\section{Scattering effect on attenuation curve} 
\label{sec:scattering_effect}
\begin{figure}
    \centering
    \includegraphics[width =\linewidth, clip]{\figdir/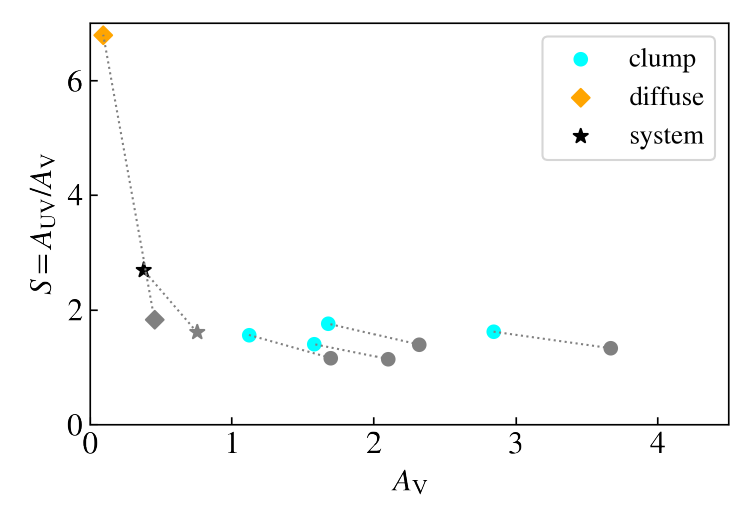}
    \caption{$A_{\rm V}$ vs Slope relation for FL957 at
$z=7.70$. The cyan, orange, and black symbols
show the clump, diffuse, and system-integrated
components, respectively, including scattering
effects. The corresponding gray symbols show
the same components without scattering. The
dotted lines connect each component with and
without scattering.}
    \label{fig:Av_S_wo_scattering}
\end{figure}

Using the \texttt{primarydirect.fits} output from SKIRT, we can
compute the $A_{\rm V}$--$S$ relation excluding scattering effects. Figure \ref{fig:Av_S_wo_scattering} shows the $A_{\rm V}-S$
relation for FL957 at $z=7.70$ for each component with scattering (colored symbols) and
without scattering (gray symbols). We find that the diffuse component in particular shows a
flatter slope ($S=1.8$) without scattering, compared to $S=6.8$ with scattering. This confirms that scattering is the dominant driver of the steep attenuation curve in the diffuse component. For the clump components (i.e., large V-band attenuation with $A_{\rm V} > 1$), the scattering effect decreases, and the corresponding change in $S$ is smaller, indicating that dust-star geometry plays a more important role there \citep[see also][]{Matsumoto:2026}.

\section{The case of SMC-like dust} \label{sec:SMC_dust}
In the main text, we investigated dust observables such as $T_{\rm d, peak}$, $A_{\rm V}$, attenuation curves, and IRX--$\Delta\beta$ assuming MW-like dust. 
However, the SMC dust model is also a commonly adopted choice in the literature for high-redshift galaxy studies, as the SMC provides a good analogue for high-redshift galaxies due to its low-metallicity environment \citep[0.15-0.2 $Z_\odot$; ][]{Choudhury:2018, Reddy:2018, Liang:2021_IRX_beta, Vijayan:2022}. Therefore, in this Appendix, we examine the case of SMC-like dust to test the robustness of our results to different dust compositions.

\begin{figure*}
    \centering
    \includegraphics[width =\linewidth, clip]{\figdir/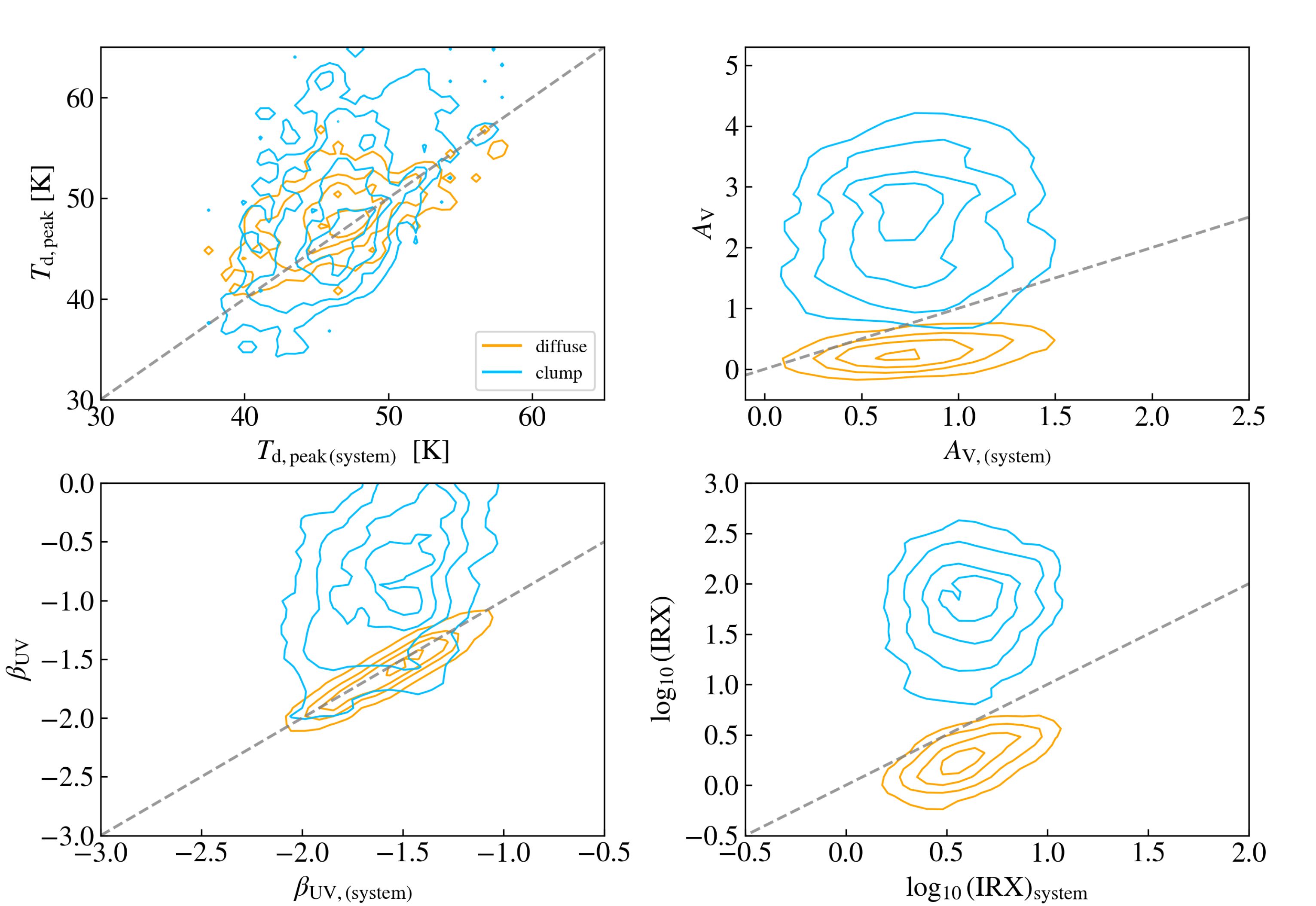}
    \vspace{-5mm}
    \caption{Comparison of dust observables between system-integrated values and individual components for all identified clumpy galaxies at $z=6-9$ as the same as Figure \ref{fig:dust_properties_statistical}.
}
    \label{fig:dust_properties_statistical_SMC}
\end{figure*}

Figure \ref{fig:dust_properties_statistical_SMC} shows a comparison of dust observables between system-integrated values and individual components (clumps and diffuse regions) for SMC-like dust. Since the SMC extinction curve is steeper than the MW curve at rest-frame UV wavelengths and lacks a UV bump, the UV slope $\beta_{\rm UV}$ is shifted redder by $\langle \beta_{\rm UV, SMC} \rangle - \langle \beta_{\rm UV, MW} \rangle \sim 0.3$--0.7 (see the bottom-left panel of Figure \ref{fig:dust_properties_statistical_SMC} and Table \ref{table:statistical_dust_properties_SMC}). However, the median values and scatter from the system-integrated values for other properties such as peak dust temperature and attenuation ($A_{\rm V}$, $A_{\rm UV}$) are similar to those of MW-like dust.

\begin{figure}
    \centering
    \includegraphics[width =\linewidth, clip]{\figdir/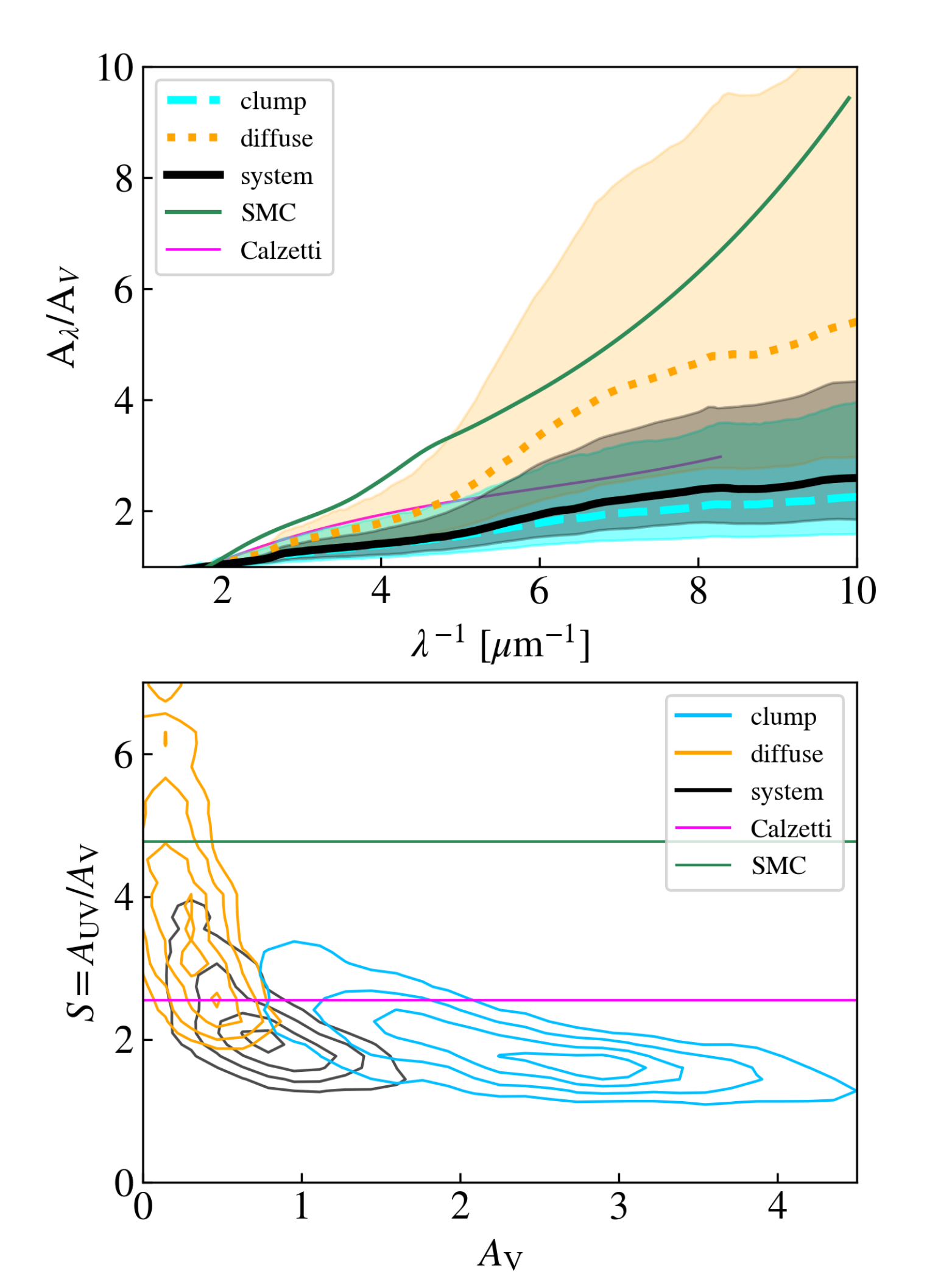}
    \vspace{-6mm}
    \caption{ Dust attenuation curves (top panel) and attenuation curve slope ($S \equiv A_{\rm UV}/A_{\rm V}$) versus V-band dust attenuation ($A_{\rm V}$) for all identified clumpy galaxies at $z=6-9$ (bottom panel) as the same as Figure \ref{fig:attenuation_curve}. The SMC extinction curve is adopted from \citet{Gordon:2003}.
}
    \label{fig:attenuation_curve_SMC}
\end{figure}

Figure \ref{fig:attenuation_curve_SMC} shows dust attenuation curves for SMC-like dust. Although the dust composition differs, the attenuation curves for system-integrated values are similar to those of MW-like dust: grayer than the Calzetti attenuation law with $\langle S_{\rm Cal} \rangle - \langle S_{\rm SMC(MW)} \rangle = 0.5\, (0.7)$. The behavior that attenuation curves for clumps (diffuse components) are grayer (steeper) than the system-integrated curves remains the same as in the MW case.

\begin{figure*}
    \centering
    \includegraphics[width =\linewidth, clip]{\figdir/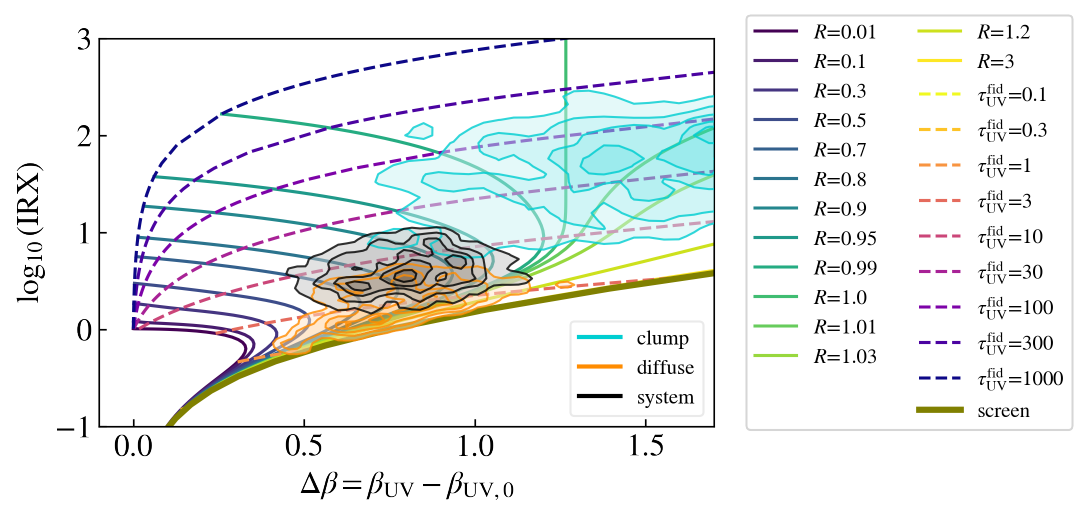}
    \vspace{-5mm}
    \caption{IRX--$\Delta\beta$ relation with toy model predictions, where $\Delta\beta \equiv \beta_{\rm UV} - \beta_{\rm UV,0}$ represents the change in UV slope due to dust attenuation, as the same as Figure \ref{fig:IRX_delta_all_MW}. }
    \label{fig:IRX_delta_all_SMC}
\end{figure*}

For the IRX--$\Delta \beta$ relations, we plot them in Figure \ref{fig:IRX_delta_all_SMC}, respectively. The toy model grids introduced in Section \ref{subsec:toy_model} are calculated using the SMC extinction curve. Since the IRX--$\Delta \beta$ relation depends only on geometry and dust column density for a fixed dust composition, the locations of clumps, diffuse components, and system-integrated values occupy the same coordinates in the $(R, \tau_{\rm UV}^{\rm fid})$ parameter space as in the MW case \footnote{Although the dust mass opacity $\kappa_{\rm UV}$ at 1600~\AA \,
differs slightly between MW and SMC dust models 
($7.45 \times 10^4$ and $8.25 \times 10^4\,{\rm cm^2\,g^{-1}}$, respectively; 
\citealt{Weingartner_Draine:2001}), this $\sim$10\% difference results in 
nearly identical positions in the $(R, \tau^{\rm fid}_{\rm UV})$ parameter 
space.}.

In summary, while the choice of dust composition (MW vs. SMC) affects the absolute values of UV slopes due to differences in extinction curve shapes, the relative differences between clumps, diffuse regions, and system-integrated values remain consistent. Our main conclusions regarding the spatial variations of dust observables and their physical interpretation via the toy model are robust to the choice of dust composition.

\begin{comment}
\begin{table*}[ht]
\centering
\renewcommand{\arraystretch}{1.2}
\begin{tabular}{lcccccccccc}
\hline
Component & number & $\langle\beta_{\rm UV}\rangle$ & $\langle\beta_{\rm UV,0}\rangle$ & $\langle \log_{10}({\rm IRX})\rangle$ 
& $\langle L_{\rm UV,1600} \rangle$ [$L_\odot$] & $\langle L_{\rm IR} \rangle $ [$L_\odot$] 
& $\langle T_{\rm d, peak} \rangle$ [K] & $\langle A_{\rm UV}\rangle$ & $\langle A_V \rangle $ & $\langle S\rangle$ \\
\hline \hline
Clump   & 1059 & -0.74 & -2.49 & 1.8 & $7.38\times10^{9}$ & $4.23\times10^{11}$ & 48.04 & 4.24 & 2.36 & 1.75 \\
Diffuse & 376  & -1.59 & -2.38 & 0.27 & $7.97\times10^{11}$ & $1.41\times10^{12}$ & 47.04 & 0.94 & 0.35 & 3.69 \\
System  & 376  & -1.56 & -2.37 & 0.59 & $ 8.12\times10^{11}$ & $3.24\times10^{12}$ & 46.52 & 1.50 & 0.76 & 1.99 \\
\hline
\end{tabular}
\caption{Summary of median dust properties for clumpy galaxies, as the same as Table \ref{table:statistical_dust_properties}.} \label{table:statistical_dust_properties_SMC}
\end{table*}
\end{comment}
\begin{table*}[ht]
\centering
\renewcommand{\arraystretch}{1.2}
\begin{tabular}{lcccccc}
\hline
Component & Number & $\langle\beta_{\rm UV}\rangle$ & $\langle\beta_{\rm UV,0}\rangle$ & $\langle \log_{10}({\rm IRX})\rangle$ 
& $\langle L_{\rm UV,1600} \rangle$ [$L_\odot$] & $\langle L_{\rm IR} \rangle$ [$L_\odot$] \\
\hline \hline
Clump   & 1059 & $-0.69_{-0.53}^{+0.55}$ & $-2.49_{-0.21}^{+0.30}$ & $1.75_{-0.41}^{+0.34}$ & $1.05_{-0.63}^{+2.16}\times 10^{9}$ & $5.21_{-3.27}^{+19.0}\times 10^{10}$ \\
Diffuse & 376  & $-1.57_{-0.24}^{+0.22}$ & $-2.36_{-0.16}^{+0.17}$ & $0.29_{-0.22}^{+0.20}$ & $1.13_{-0.39}^{+0.83}\times 10^{11}$ & $2.13_{-1.06}^{+2.59}\times10^{11}$ \\
System  & 376  & $-1.53_{-0.26}^{+0.23}$ & $-2.36_{-0.18}^{+0.17}$ & $0.60_{-0.16}^{+0.20}$ & $1.16_{-0.40}^{+0.89}\times 10^{11}$ & $4.93_{-2.29}^{+4.58}\times 10^{11}$ \\
\hline 
\noalign{\vspace{1mm}}
\hline
Component & $\langle T_{\rm d, peak} \rangle$ [K] & $\langle A_{\rm UV}\rangle$ & $\langle A_V \rangle$ & $\langle S\rangle$ \\
\hline \hline
Clump   & $48.59_{-5.6}^{+6.5}$ & $4.10_{-1.00}^{+0.82}$ & $2.17_{-0.81}^{+0.87}$ & $1.85_{-0.30}^{+0.59}$ \\
Diffuse & $48.22_{-2.5}^{+2.5}$ & $0.96_{-0.28}^{+0.26}$ & $0.27_{-0.12}^{+0.19}$ & $3.59_{-1.00}^{+1.37}$ \\
System  & $46.51_{-2.9}^{+2.4}$ & $1.53_{-0.30}^{+0.35}$ & $0.78_{-0.25}^{+0.26}$ & $2.00_{-0.29}^{+0.43}$ \\
\hline
\end{tabular}
\caption{Summary of median dust observables for clumpy galaxies, as the same as Table \ref{table:statistical_dust_properties}.}
\label{table:statistical_dust_properties_SMC}
\end{table*}

\begin{comment}
\begin{figure}
    \centering
    \includegraphics[width =\linewidth, clip]{\figdir/IRX_beta_all_SMC.pdf}
    \vspace{-6mm}
    \caption{IRX--$\beta_{\rm UV}$ relation for all identified clumpy galaxies at $z=6$--9 as the same as Figure \ref{fig:IRX_beta}.}
    \label{fig:IRX_beta_SMC}
\end{figure}

\begin{figure*}
    \centering
    \includegraphics[width =\linewidth, clip]{\figdir/IRX_beta_IRX_delta_beta_all_SMC_w_toy_model.pdf}
    \vspace{-5mm}
    \caption{IRX--$\beta_{\rm UV}$ relation (left) and IRX--$\Delta\beta$ relation (right) with toy model predictions, where $\Delta\beta \equiv \beta_{\rm UV} - \beta_{\rm UV,0}$ represents the change in UV slope due to dust attenuation, as the same as Figure \ref{fig:IRX_delta_all_MW}. }
    \label{fig:IRX_delta_all_SMC}
\end{figure*}
\end{comment}

\bibliography{FL_dust}{}
\bibliographystyle{aasjournal}

\end{document}